\documentclass[usenatbib,useAMS,usedcolumn]{mn2e}
\usepackage{longtable}
\usepackage{psfig}
\usepackage{epsfig}
\usepackage{graphicx}
\usepackage{times}
\usepackage{euscript}
\def\arcm{\hbox{$^\prime$}}

\def\deg{\hbox{$^\circ$}}
\def\spose#1{\hbox to 0pt{#1\hss}}
\def\gtsim{$\mathrel{\spose{\lower 3pt\hbox{$\sim$}}
        \raise 2.0pt\hbox{$>$}}$\thinspace}
\def\simpropto{$\mathrel{\spose{\lower 3pt\hbox{$\sim$}}
        \raise 2.0pt\hbox{$\propto$}}$\thinspace}
\newcommand{\rosat}{\emph{ROSAT}}
\newcommand{\chandra}{\emph{Chandra}}
\newcommand{\xmm}{\emph{XMM-Newton}}
\newcommand{\asca}{\emph{ASCA}}
\newcommand{\einstein}{\emph{Einstein}}
\newcommand{\arcs}{\mbox{\arcm\hskip -0.1em\arcm}}
\newcommand{\Lx}{\ensuremath{L_{\mathrm{X}}}}

\newcommand{\Zsol}{\ensuremath{Z_{\odot}}}

\newcommand{\Msol}{\ensuremath{M_{\odot}}}

\newcommand{\LBsol}{\ensuremath{L_{B\odot}}}

\newcommand{\Bfit}{\ensuremath{\beta_{fit}}}

\newcommand{\NH}{\ensuremath{N_{\mathrm{H}}}}

\newcommand{\s}{\ensuremath{\mbox{~s}}}
\newcommand{\ps}{\ensuremath{\s^{-1}}}
\newcommand{\km}{\ensuremath{\mbox{~km}}}
\newcommand{\Mpc}{\ensuremath{\mbox{~Mpc}}}
\newcommand{\pMpc}{\ensuremath{\Mpc^{-1}}}
\newcommand{\kmpspMpc}{\ensuremath{\km \ps \pMpc\,}}
\newcommand{\erg}{\ensuremath{\mbox{~erg}}}
\newcommand{\ergps}{\ensuremath{\erg \ps}}
\newcommand{\kmps}{\ensuremath{\km \ps}}
\newcommand{\sas}{\textsc{sas}}
\newcommand{\Ho}{\ensuremath{H_\mathrm{0}}}

\newcommand{\Rth}{\ensuremath{R_{\mathrm{200}}}}
\newcommand{\Dtf}{\ensuremath{D_{\mathrm{25}}}}

\newcommand{\hs}{\ensuremath{h_{\mathrm{75}}}}
\begin{document}

\title[
AWM 4 - an isothermal cluster observed with XMM-Newton
]
{
AWM 4 - an isothermal cluster observed with XMM-Newton
}
\author[
E. O'Sullivan et al. 
]
{
E. O'Sullivan\footnotemark$^{1}$, J. M. Vrtilek$^{1}$, J.~C. Kempner$^{1,2}$,
L.~P. David$^{1}$ , J.~C. Houck$^{3}$\\
$^{1}$ Harvard Smithsonian Center for Astrophysics,
60 Garden Street, Cambridge, MA 02138, USA\\
$^{2}$ Dept. of Physics and Astronomy, Bowdoin College, 8800 College Station, Brunswick, ME 04011, USA \\
$^{3}$ MIT Center for Space Research, 77 Massachusetts Avenue, Cambridge,
MA 02139, USA \\ 
\\
}

\date{Accepted 2003 ?? Received 2003 ??; in original form 2003 ??}
\pagerange{\pageref{firstpage}--\pageref{lastpage}}
\def\LaTeX{L\kern-.36em\raise.3ex\hbox{a}\kern-.15em
    T\kern-.1667em\lower.7ex\hbox{E}\kern-.125emX}

\label{firstpage}

\maketitle

\begin{abstract}
  We present analysis of an \xmm\ observation of the poor cluster AWM~4.
  The cluster is relaxed and its X-ray halo is regular with no apparent
  substructure. Azimuthally averaged radial spectral profiles suggest that
  the cluster is isothermal to a radius of at least 160~kpc, with no
  evidence of a central cooling region. Spectral mapping shows some
  significant temperature and abundance substructure, but no evidence of
  strong cooling in the cluster core. Abundance increases in the core, but
  not to the extent expected, and we find some indication of gas mixing.
  Modeling the three dimensional properties of the system, we show that
  ongoing heating by an AGN in the dominant elliptical, NGC~6051, is likely
  to be responsible for the lack of cooling. We also compare AWM~4 to
  MKW~4, a cluster of similar mass observed recently with \xmm. While the
  two systems have similar gravitational mass profiles, MKW~4 has a cool
  core and somewhat steeper gas density profile, which leads to a lower
  core entropy. AWM~4 has a considerably larger gas fraction at
  0.1$\times$\Rth, and we show that these differences result from the
  difference in mass between the two dominant galaxies and the activity
  cycles of their AGN. We estimate the energy required to raise the
  temperature profile of MKW~4 to match that of AWM~4 to be
  9$\times$10$^{58}$ erg, or 3$\times$10$^{43}$ \ergps\ for 100 Myr,
  comparable to the likely power output of the AGN in AWM~4.
\end{abstract}


\footnotetext{Email: ejos@head.cfa.harvard.edu}

\section{Introduction}
\label{sec-intro}
AWM~4 is a poor cluster consisting of $\sim$30 galaxies centred on the
giant elliptical NGC~6051 \citep{KoranyiGeller02}. The association of
galaxies was first identified in the survey of \citet{Albertetal77}, and
the presence of a deep potential well was later confirmed by X--ray
observations \citep{Krissetal83}. Of the identified galaxies in the group,
most are absorption line systems, with a strong concentration of early-type
galaxies toward the core \citep{KoranyiGeller02}. The group is apparently
relaxed, with a smooth velocity distribution about a mean redshift of
$cz$=9520 \kmps. In conjunction with the morphology segregation, this suggests
that it has been undisturbed by major interactions for some time.

The dominant elliptical in the cluster is considerably more luminous than
the galaxies around it, with a difference in magnitude above the
second-ranked galaxy of $M_{12}$=1.426. NGC~6051 is well described by an
r$^{\frac{1}{4}}$ law surface brightness profile, and does not show signs
of an extended stellar envelope \citep{Schombert87}. NGC~6051 has a
powerful active nucleus (4C +24.36), with
radio lobes extending out roughly along the minor axis of the galaxy. There
is no sign of a central point source at other wavelengths, suggesting that
the axis of the AGN jets is in the plane of the sky, and that the central
engine is highly absorbed. The galaxy major axis is roughly aligned with
that of the galaxy distribution, and that of the X--ray halo.
Table~\ref{tab-props} gives details of the position and scale of the
cluster and its dominant galaxy.

The cluster has been observed in the X-ray band previously, by the
\einstein, \asca\ and \rosat\ observatories. These observations have shown
it to be extended, regular, and relatively luminous. The temperature of the
halo gas has been shown to be $\sim$2.5 keV
\citep{Krissetal83,Finoguenovetal01}, and several estimates of the mass of
the system have been made. These agree reasonably well with the mass
estimated from the galaxy distribution, $\sim$9$\times$10$^{13}$ \Msol\ 
within 600 kpc \citep{KoranyiGeller02}.  Analysis of the \asca\ data for
the cluster shows it to have a relatively flat temperature profile, though
the poor spatial resolution of the instrument prevents the identification
of any small scale cooling in the core of the cluster.

AWM~4 lies at the low end of the mass and temperature range usually
associated with clusters.  While more massive clusters have been relatively
thoroughly studied and are currently receiving much attention, the lower
luminosities of galaxy groups have made the exploration of their properties
more difficult. Systems at the border between these two classes are
particularly interesting, in that they may shed light on the mechanisms
behind the differences between clusters and groups. AWM~4 is also of
interest because of the large radio source associated with NGC~6051.
Several \chandra\ studies in larger systems have found evidence that AGN
jets can evacuate cavities in the hot intra-cluster medium (ICM), and may
have a part in regulating cooling \citep[e.g.,][]{Fabianetal00,McNamara00}.
The ICM in AWM~4 is cooler than in these more massive systems, and this
gives us an opportunity to observe the effects of a powerful AGN on a
relatively small cluster.

In this paper, we use a recent \xmm\ observation of AWM~4 to study the
structure of the cluster halo. Section~\ref{sec-obs} describes the
observation and the reduction of the data, and Section~\ref{sec-results}
describes our data analysis and our main results. We discuss these in
Section~\ref{sec-discuss}, and compare them with results for other systems.
Our conclusions are given in Section~\ref{sec-conc}. Throughout the paper
we assume \Ho=75 \kmpspMpc\ and normalise optical luminosities to the
B-band luminosity of the sun, \LBsol=5.2$\times$10$^{32}$ \ergps.
Abundances are measured relative to the abundance ratios of
\citet{AndersGrevesse79}.  These differ from the more recent ratios given
by \citet{GrevesseSauval98} in that the abundance of Fe is a factor of
$\sim$1.4 lower. This means that our measured abundances are underestimated
by a factor of $\sim$1.4 when compared with those estimated using the more
recent ratios. This should be noted when making comparisons with other
results.

\begin{table}
\begin{center}
\begin{tabular}{lc}
\hline
R.A. (J2000) & 16 04 57.0 \\
Dec. (J2000) & +23 55 14 \\
Redshift & 9520 \kmps\ \\
Distance (\Ho=75) & 126.9 Mpc \\
1 arcmin = & 36.9 kpc \\
NGC~6051 \Dtf\ radius & 28.6 kpc\\
\hline
\end{tabular}
\end{center}
\caption{\label{tab-props} Location and scale for AWM 4.}
\end{table}

\section{Observation and Data Reduction}
\label{sec-obs}
AWM~4 was observed with \xmm\ during orbit 573 (2003 January 25-26) for
just over 20,000 seconds. The EPIC MOS and PN instruments were operated in
full frame and extended full frame modes respectively, with the medium
filter. A detailed summary of the \xmm\ mission and instrumentation can be
found in \citet[and references therein]{Jansenetal01}. The raw data from
the EPIC instruments were processed with the most recent version of
the \xmm\ Science Analysis System (\textsc{sas v.5.4.1}), using the
\textsc{epchain} and \textsc{emchain} tasks. After filtering for bad pixels
and columns, X--ray events corresponding to patterns 0-12 for the two MOS
cameras and patterns 0-4 for the PN camera were accepted.  Investigation of
the total count rate for the field revealed a short background flare in the
second half of the observation. Times when the total count rate deviated
from the mean by more than 3$\sigma$ were therefore excluded. The effective
exposure times for the MOS 1, MOS 2 and PN cameras were 17.3, 17.4 and 12.7
ksec respectively.

Images and spectra were extracted from the cleaned events lists with the
\sas\ task \textsc{evselect}. For simple imaging analysis, the filtering
described above was considered sufficient. Event sets for use in spectral
analysis were further cleaned using \textsc{evselect} with the expression
`(FLAG == 0)' to remove all events potentially contaminated by bad pixels
and columns. All data within 17\arcs\ of point sources were also removed,
excluding the false source detection for the core of NGC 6051. We allowed
the use of both single and double events in the PN spectra, and single,
double, triple and quadruple events in the MOS spectra. Response files were
generated using the \sas\ tasks \textsc{rmfgen} and \textsc{arfgen}. 

\begin{table}
\begin{center}
\begin{tabular}{l|ccc}[!h]
Dataset & MOS 1 & MOS 2 & PN \\
\hline
Blank field & 0.0312 & 0.0282 & 0.121 \\
Telescope closed & 0.142 & 0.172 & 0.609 \\
\end{tabular}
\end{center}
\caption{\label{tab-scaling} Scaling factors used in creation of background
  spectra and images. The factors show the scaling between the source
  dataset and the blank field or telescope closed data for each
  instrument.}
\end{table}

Background images and spectra were generated using the ``double
subtraction'' method \citep{Arnaudetal02,Prattetal01}. The ``blank field''
background data sets of \citet{ReadPonman03} and the ``telescope closed'',
particles only data sets of \citet{Martyetal02} form the basis of this
process. Background images are generated by scaling the ``closed'' data to
match the measured events outside the telescope field of view. The
``blank'' data are then scaled to match the observation exposure time, and
a ``soft excess'' spectrum calculated by comparison of this scaled
background to the low energy observed spectrum in the outer part of the
detector field of view. The source does not extend to the outer part of the
field of view, so the soft excess spectrum should measure only the
difference in background soft emission between the ``blank'' data and that
for the target. The various background components can then be combined to
form the background images. A similar process is used to create background
spectra, again scaling the ``blank'' data to match the observation, and
correcting for differences in the soft background component using a large
radius spectrum. Scaling factors used to correct the ``blank'' and
``closed' data are shown in table~\ref{tab-scaling}. The ``soft excess''
spectrum for the PN resembles a 0.6 keV Bremsstrahlung model, but while the
MOS spectra are consistent with this they are of poorer quality and contain
elements which appear to be associated with spectral lines in the
background, most notably the 1.55 keV Al-K fluorescent line. Neither PN nor
MOS ``soft excess'' spectra can be accurately represented by a simple
model, as is to be expected given the method by which they are constructed.
However, the apparent low temperature of the spectrum confirms that we are
not including cluster emission in the background, as \rosat\ spectra
suggest that at the radius used, the cluster temperature is $\sim$1.6 keV
\citep{Helsdonponman00}.  The normalisation of the Bremsstrahlung model is
$\sim$7.6$\times 10^{-8}$ arcmin$^{-2}$ for the PN and $\sim$1.1 $\times
10^{-8}$ arcmin$^{-2}$ for the MOS.

In some cases, where the signal-to-noise of our spectra was
relatively low, we tested the quality of these ``double'' background
spectra to determine whether we might be introducing error through the
scaling process. We selected an annular region in the observation data set,
at large radius where the source contribution should be negligible, and
used it as an alternate background. We then compared the resulting spectral
fits. In all cases the fits were similar, with no qualitative difference
introduced by the use of the ``double'' background spectra. We therefore
consider these to be an accurate measure of the background in all cases.

%%%FIGS FROM RESULTS SECTION>
\begin{figure}
\centerline{\epsfig{width=8.5cm,file=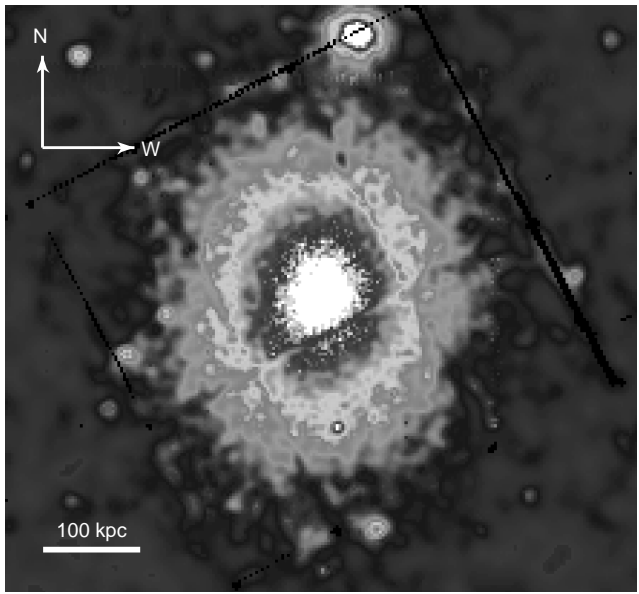}}
\caption{
\label{fig-xray} Adaptively smoothed X--ray image of AWM~4, combining data
from the PN and MOS cameras.
}
\end{figure}

\begin{figure}
\centerline{\epsfig{width=8.5cm,file=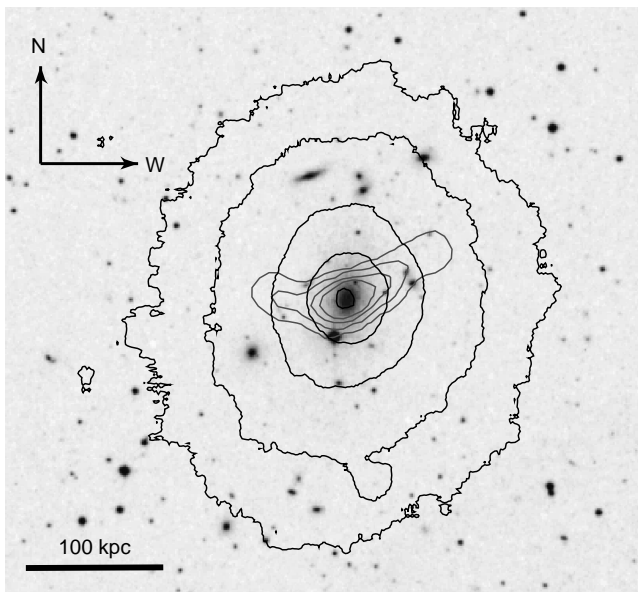}}
\caption{
\label{fig-ovly} Digitized Sky Survey image of the core of AWM~4 with
X--ray and radio contours superimposed. X--ray contours (black) are taken
from our MOS images, while the radio contours (shown in grey) are from the
VLA First 20cm survey. Higher resolution radio contours can be seen in Figure~\ref{fig-Tmap}.}
\end{figure}

\section{Results}
\label{sec-results}
Figure~\ref{fig-xray} shows an exposure and vignetting corrected mosaiced
image of AWM~4, combining data from the PN and two MOS cameras. The image
was smoothed using the \textsc{sas} task \textsc{asmooth} with smoothing
scales chosen to achieve a signal-to-noise ratio of 10. The X-ray halo is
highly extended and elliptical, with the major axis running approximately
north-south. Ignoring the distortion introduced by the PN chip gaps, the
cluster appears quite regular, with no major substructure. Several
point-like sources are visible in the field, as well as the bright
background Seyfert 1 galaxy J160452.8+240235 \citep{Veronetal01}, visible
to the north of the cluster near a chip gap. The extent of the radio source
associated with NGC~6051 can be seen in Figure~\ref{fig-ovly}. The two
lobes extend well beyond the central galaxy, to at least 70~kpc eastward
and 85~kpc to the northwest.

%%%FIGS MOVED UP...

\subsection{Two-dimensional surface brightness modeling}
We created images in the 0.5-3.0 keV band for each detector, with
associated background images, and exposure maps. We used the \sas\ task
\textsc{calview} to produce on-axis PSF images for each camera in this
energy band. All images were binned to 4.4\arcs\ pixels. We then
simultaneously fit these images using \textsc{sherpa} (running in
\textsc{ciao} v2.3). As the data contain numerous pixels which have small
numbers of (or zero) counts, we used the Cash statistic \citep{Cash79}.
This requires us to model the background, and we use flat models, with the
normalisation free to vary independently for each camera. This will
introduce some inaccuracies, as the background is not flat. However, the
variation relative to the source is small, and using a background model
allows us effectively to ignore small variations in the background data
sets. The Cash statistic does not provide an absolute measure of the
goodness of the fit, so we used the residual images and azimuthally
averaged radial profiles to check the fit quality. 

\begin{figure*}
\centerline{
\mbox{
\epsfig{file=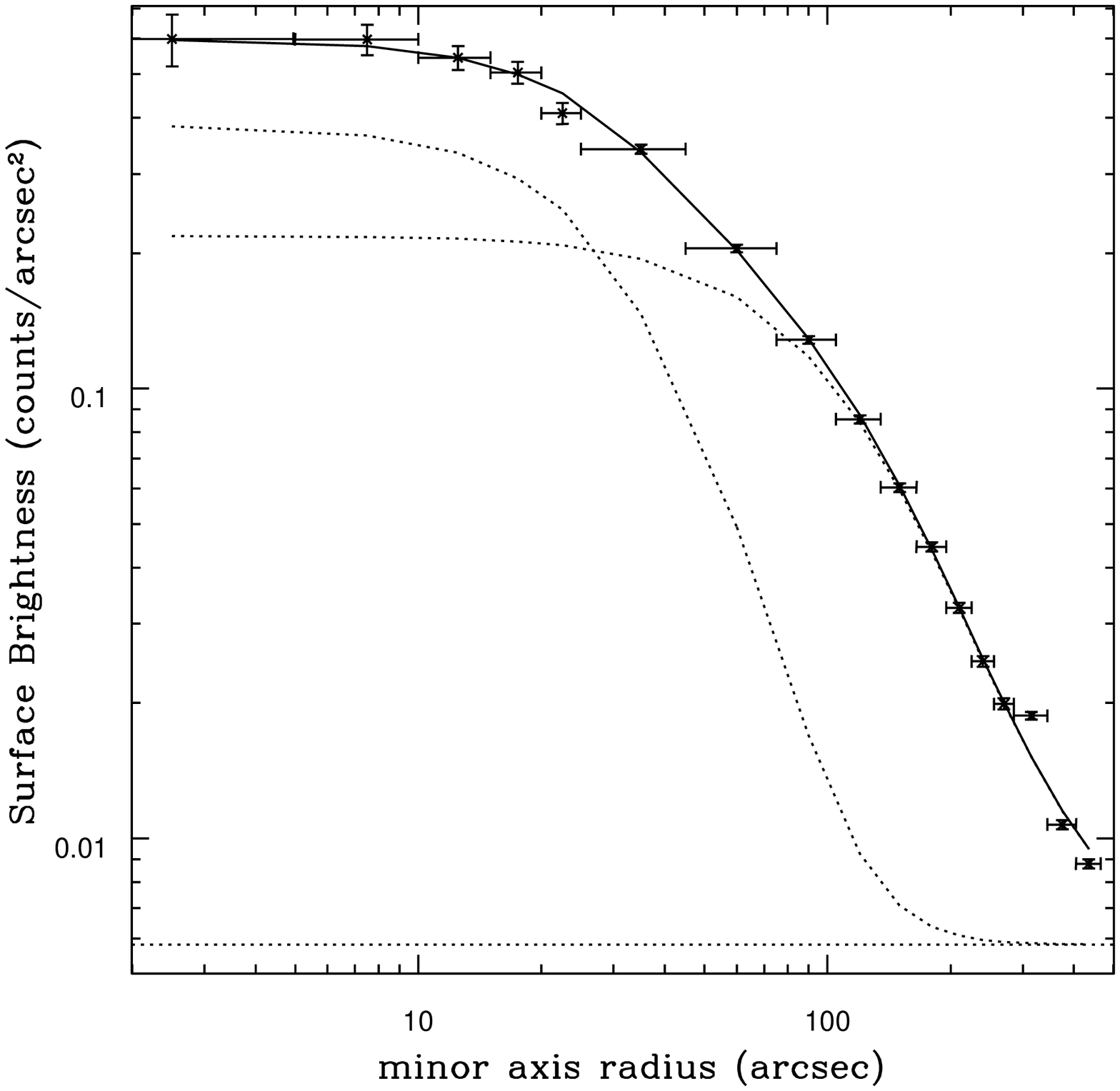,width=9cm,bbllx=20,bblly=200,bburx=592,bbury=779,clip=}
\epsfig{file=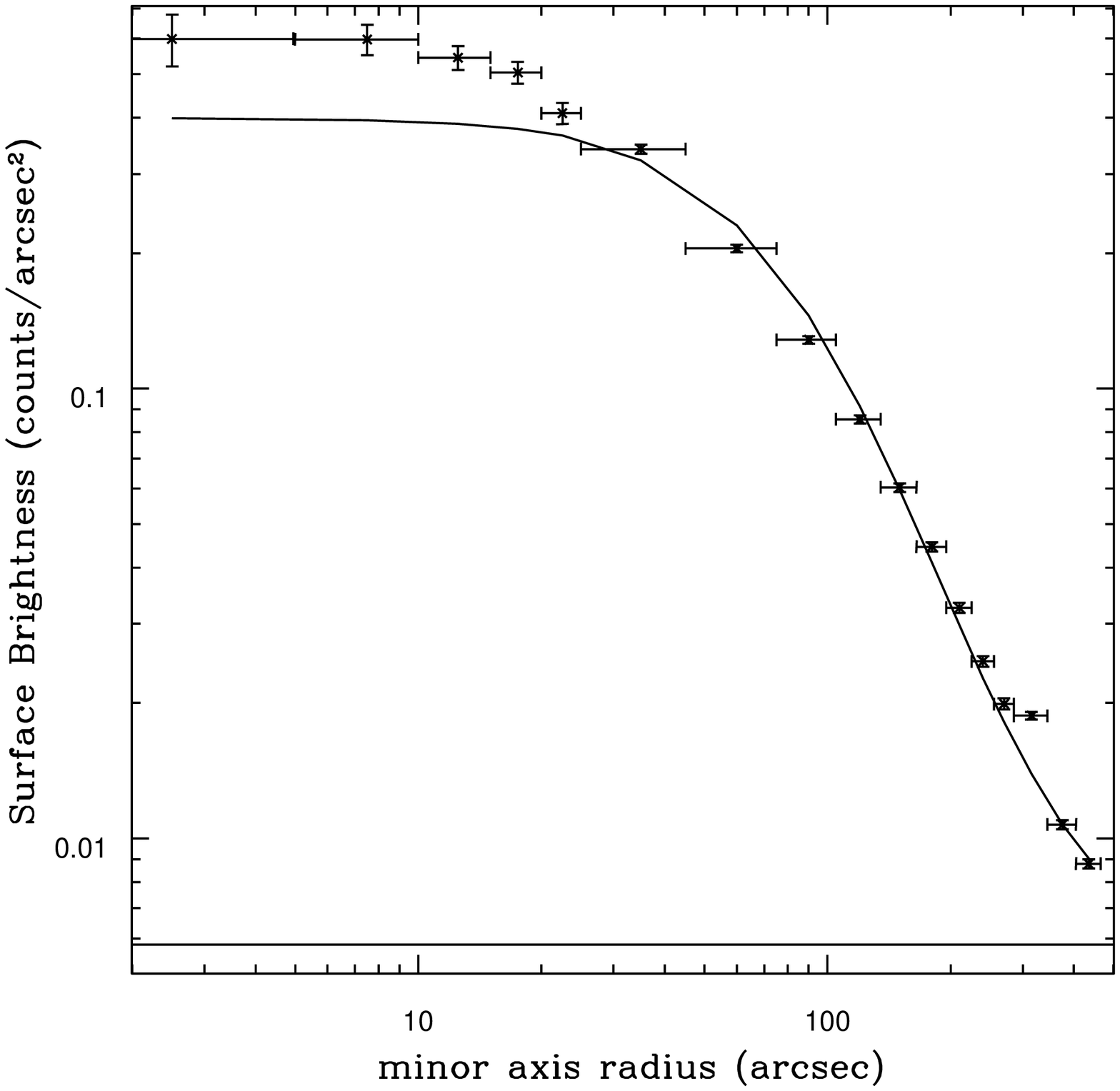,width=9cm,bbllx=20,bblly=200,bburx=592,bbury=779,clip=}
}}
\caption{
\label{fig-SB} Radial surface brightness profiles for the 2-component
(\textit{left}) and 1-component (\textit{right}) models. In both cases the
data, which for the purpose of the profile is taken from the MOS 1 image,
is marked by black points with (1$\sigma$) error bars. In the 1-component
plot the fitted beta model and flat background model are marked by solid
lines. In the 2-component plot the combined model is marked by a solid
line, while the model components and the background are marked by dashed
lines. Note that each model component has
the background added to it in this plot, hence the inner beta model curves
to meet the background at large radii. 
}
\end{figure*}

We initially fit a single beta model, which was reasonably successful, and
allowed us to determine the position angle and ellipticity of the group
halo. The background models were allowed to vary freely during fitting.
However, it was clear from the radial profile that the fit did not
successfully model the inner part of the halo, and we therefore introduced a
second beta model, which we constrained to be circular. This produced a
satisfactory model of the halo, with no major residuals. Parameters of
these fits are given in Table~\ref{tab-SB}. Radial surface brightness
profiles for the two fits are shown in Figure~\ref{fig-SB}.

%%%%%%%%%%%%%FIGURE MOVED %%%%%%
\begin{table*}
\begin{center}
\begin{tabular}{lcccccccc}
Fit & r$_{core}$ & $\beta_{fit}$ & axis ratio & position angle & amplitude & r$_{core,2}$
& $\beta_{fit,2}$ & amplitude \\
 & (kpc) & & & (degrees) & & (kpc) & & \\
\hline\\[-2mm]
1 component & 67.32$^{+0.29}_{-2.42}$ & 0.672$^{+0.011}_{-0.002}$ &
1.20$\pm$0.01 & 174.56$^{+0.93}_{-0.94}$ & 7.702 & - & - & - \\
 & & & & & & & \\[-2mm]
2 component & 96.74$^{+19.13}_{-1.66}$ & 0.708$^{+0.054}_{-0.005}$ &
1.22$^{+0.04}_{-0.02}$ & 174.48$^{+2.39}_{-2.01}$ & 4.155 &
33.58$^{+20.58}_{-9.03}$ & 0.998$^{+0.749}_{-0.347}$ & 7.657 \\
\end{tabular}
\end{center}
\caption{
\label{tab-SB} Parameters with 1$\sigma$ errors for the best fitting
surface brightness models. Note that the 1-component fit, while adequate at
high radii, is a poor fit in the core. Errors for each model were estimated
with all parameters shown in the table free for perturbation. Position
angle is measured anti-clockwise from north.
}
\end{table*}    

Our models show a somewhat steeper profile than has previously been
obtained assuming azimuthal symmetry. Fits to \einstein\ IPC data using a
single beta model gave \Bfit$\sim$0.6 and a core radius of $\sim$130 kpc
\citep{JonesForman99}, for a region of interest $\sim$960\arcs\ (590 kpc)
in radius. Using \rosat\ PSPC data, \citet{Finoguenovetal01} found a
similar slope, \Bfit=0.62$\pm$0.02, and a rather smaller core radius,
r$_c$=68.7$\pm$2.8 kpc. However, a small region in the center of the group
was excluded in the \rosat\ analysis, in order to avoid biases associated
with a central cooling region. It seems possible that the differences
between our fits and those from \einstein\ and \rosat\ might arise from the
poorer spatial resolution of the instruments used, the removal of the core
of the group, and the differences in size of field used in the fits.

\subsection{Spectral modeling}
\label{sec-spectral}
\begin{figure}
\centerline{\epsfig{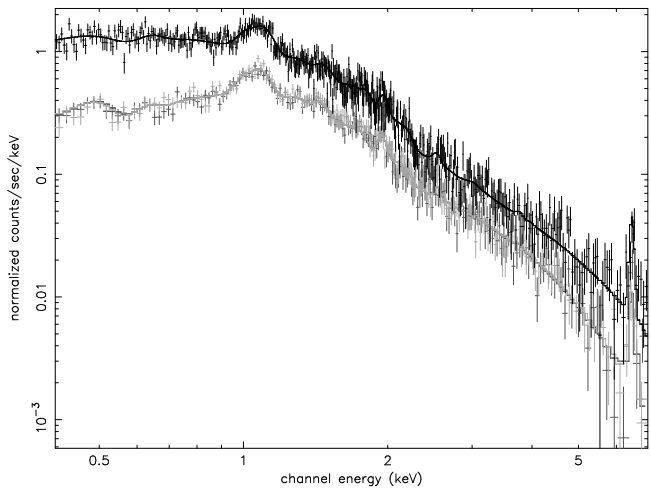}}
\caption{
\label{fig-totalspec}EPIC PN (upper, black), MOS1 and MOS 2 (lower, grey)
spectra of AWM~4, as used in our overall spectral fit to the cluster.  }
\end{figure}

Given the estimate of the ellipticity and angle of the group halo, we were
able to choose a region from which to extract an overall spectrum. We chose
a region with minor axis 260\arcs\ (158 kpc) and ellipse parameters as in the
1-component surface brightness fit. We then extracted spectra for this
region from all three cameras, and generated appropriate background
files. The resulting spectra were simultaneously fitted using \textsc{xspec}
(v11.2.0ab). We ignored energies less than 0.4 keV and greater than 8.0
keV, where the calibration of the EPIC instruments is known to be
questionable. An example of the spectra and fitted model can be seen in
Figure~\ref{fig-totalspec}.

The three spectra were adequately modeled by both the MEKAL
\citep{Liedahletal95,Kaastramewe93} and APEC \citep{Smithetal01} hot plasma
codes, leading us to believe that the X-ray emission of the group is
dominated by the contribution of the gaseous halo. Multi-temperature
models were not required, suggesting that the halo is roughly
isothermal. \NH\ was held at the Galactic value (5$\times$10$^{20}$
cm$^2$) in these fits. Details of the best fitting models are given in
Table~\ref{tab-totalspec}. The agreement between the APEC and MEKAL models
is good, and the abundances of most elements are quite comparable. One
noteworthy point is the difference in abundance between Si and Fe, with Si
being more abundant than Fe at $>$90\% significance. We note however that
this would be affected were we measure abundance relative to the ratios of
\citep{GrevesseSauval98}, as discussed in Section~\ref{sec-intro}.

\begin{table}
\begin{center}
\begin{tabular}{lcc}
Parameter & MEKAL & APEC \\
\hline\\[-3mm]
kT & 2.56$\pm$0.06 & 2.55$^{+0.05}_{-0.06}$ \\[+1mm]
Z$_{avg}$ & 0.33$^{+0.14}_{-0.11}$ & 0.47$^{+0.16}_{-0.14}$ \\[+1mm]
O & 0.16$^{+0.12}_{-0.11}$ & 0.18$^{+0.15}_{-0.13}$ \\[+1mm]
Si & 0.54$^{+0.11}_{-0.10}$ & 0.57$\pm$0.11 \\[+1mm]
S & 0.26$^{+0.13}_{-0.12}$ & 0.27$\pm$0.13 \\[+1mm]
Fe & 0.35$^{+0.03}_{-0.04}$ & 0.36$\pm$0.04 \\
\end{tabular}
\end{center}
\caption{
\label{tab-totalspec} Best fit parameters for spectral fits to the integrated
spectrum of AWM~4, extracted from a region of minor axis 260\arcs\ (158
kpc), major axis 312\arcs\ (190 kpc) and ellipse parameters taken from the
1-component surface brightness model. The ranges shown are the 90\% error
bounds for each parameter. Temperature (kT) is in keV, and all
metal abundances are relative to solar. Z$_{avg}$ is the average abundance
of all metals not listed individually.
}
\end{table}

%%%%%Fig moved down a bit...

We estimate the total luminosity within the region of interest in three
energy bands from the best fit MEKAL model. The luminosities are Log
\Lx=42.96 \ergps\ (0.5-2 keV), Log \Lx=43.22 \ergps\ (0.4-8 keV, the energy
band used in fitting) and Log \Lx=43.31 \ergps\ (0.1-12 keV). The
relatively small change between the last two values suggest that the 0.1-12
keV luminosity is a good approximation of the bolometric X-ray luminosity.

\subsubsection{Radial profiles}
\label{sec-profiles}
\begin{figure*}
\parbox[t]{1.0\textwidth}{\vspace{-1em}\includegraphics[width=9cm]{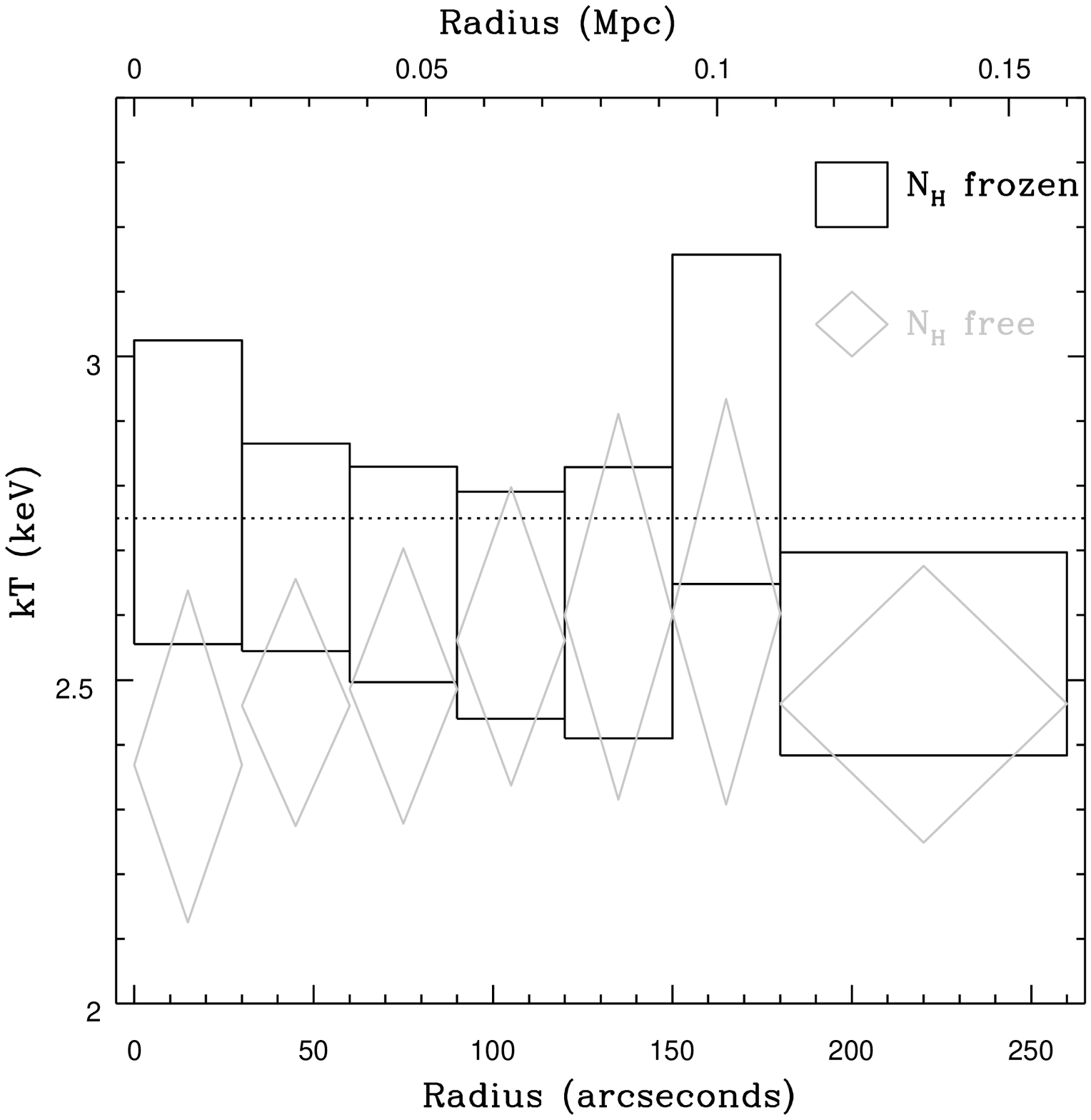}}
\parbox[t]{0.0\textwidth}{\vspace{-12.05cm}\includegraphics[width=9cm]{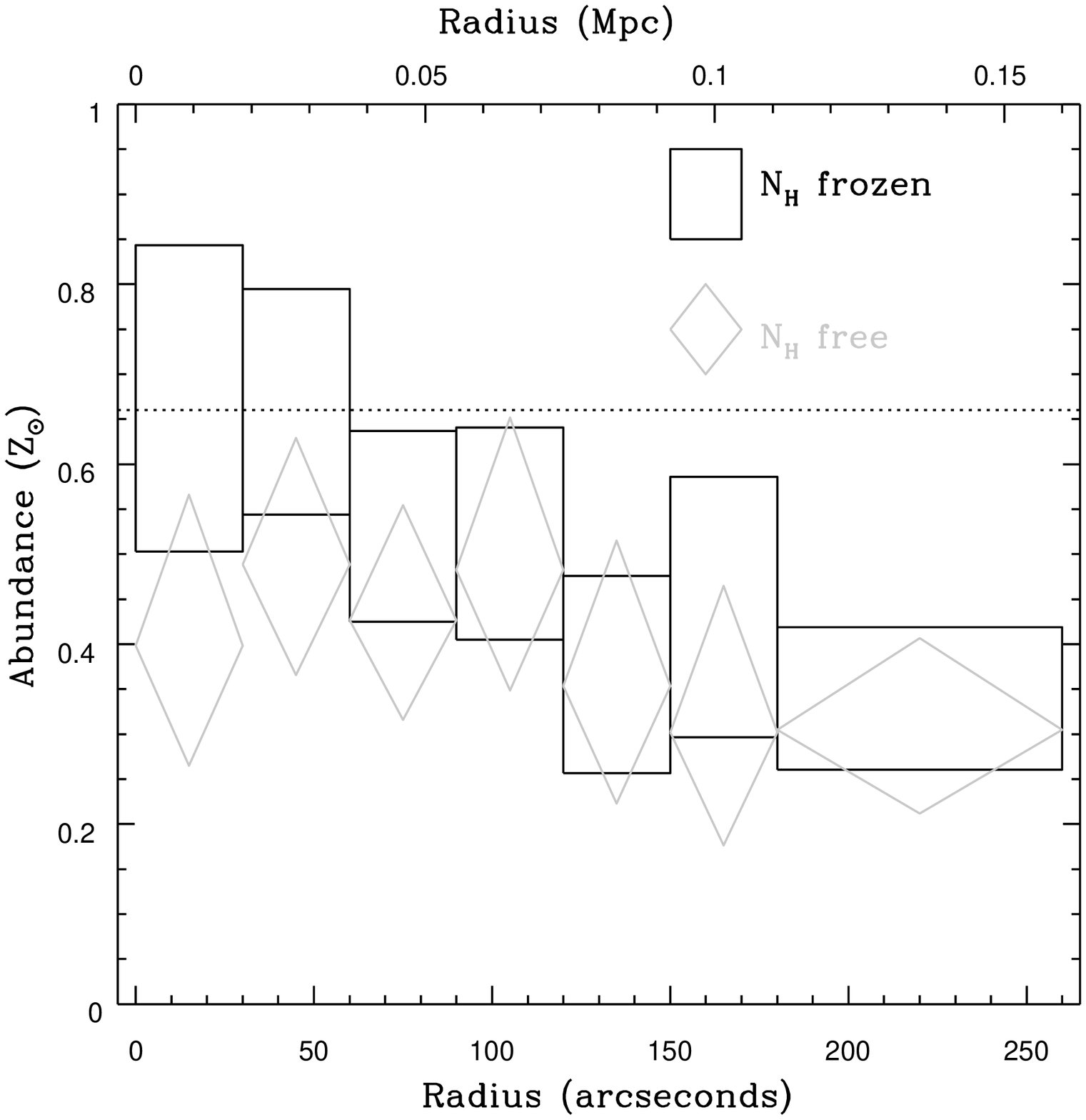}}
\parbox[t]{1.0\textwidth}{\vspace{-3.8cm}\includegraphics[width=9cm]{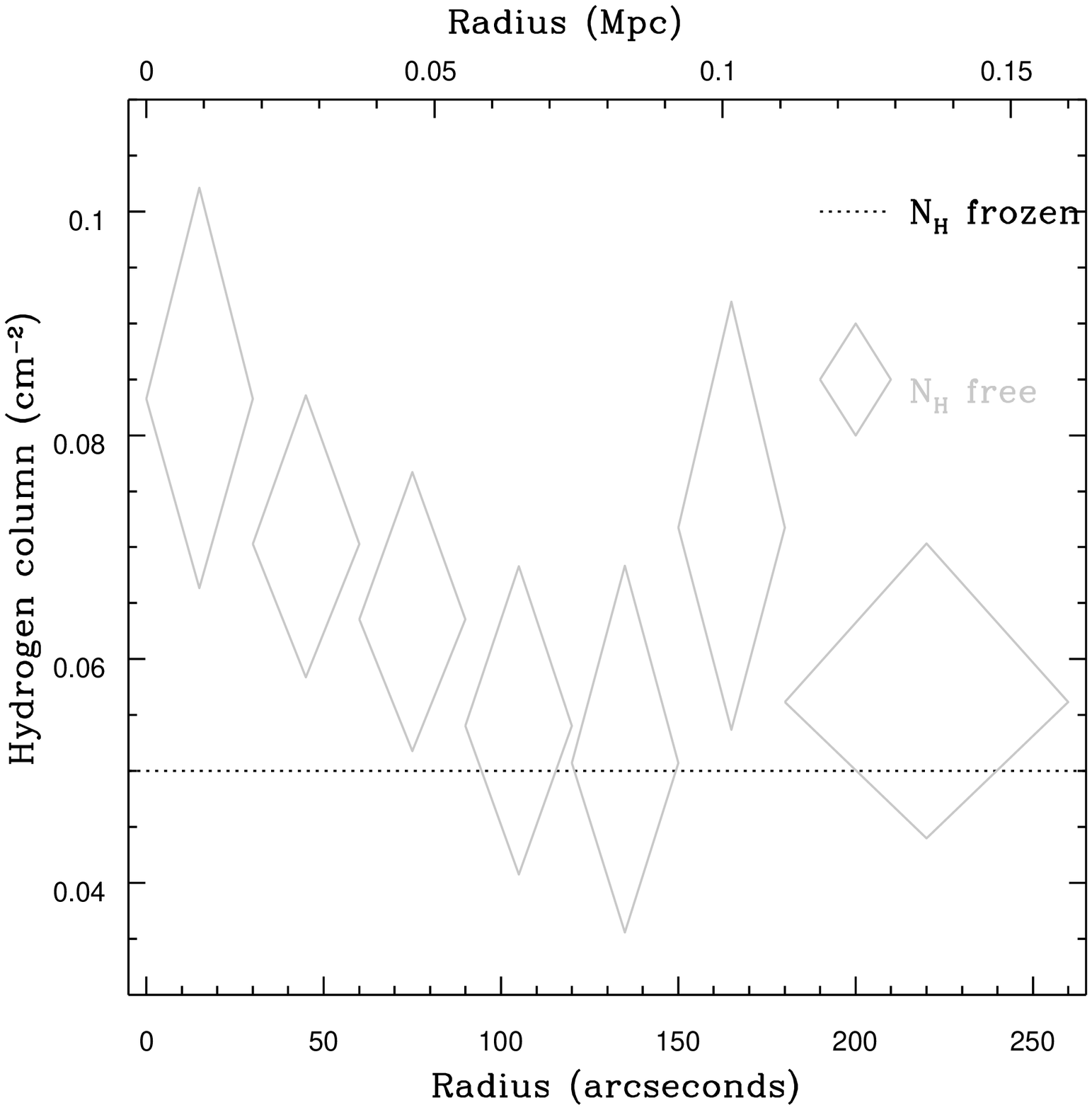}}
\parbox[t]{0.0\textwidth}{\vspace{-12.05cm}\includegraphics[width=9cm]{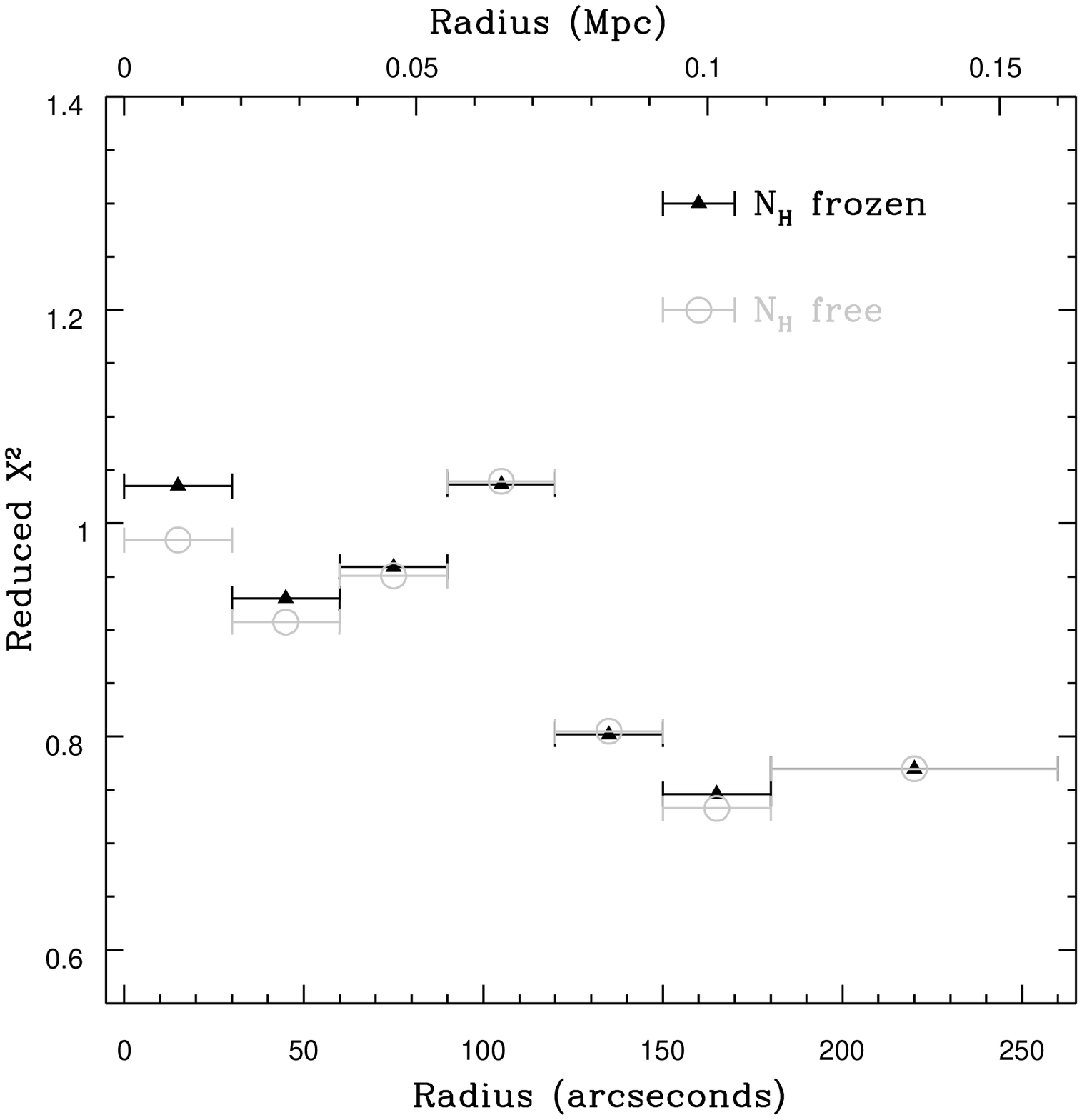}}
\vspace{-3cm}
\caption{
\label{fig-Tprof} Radial profiles of projected temperature, abundance, hydrogen
column and fit statistic (reduced $\chi_\nu^2$) for AWM~4. In each plot,
numbers along the upper axis show radius in Mpc, while those on the lower
axis show radius in arcseconds. All radii are for the minor axis of the
elliptical region. Grey symbols refer to fits carried out with hydrogen
column free to vary, black symbols to those where column is held at the
Galactic value. All regions denote 90\% errors. The dotted line in the
\NH\ plot shows the Galactic column value. The dotted lines in the kT and
abundance plots show the approximate values of these parameters measured by
\asca\ \protect\citep{Finoguenovetal01}.
}
\end{figure*}

Using elliptical annuli, we generated spectra in radial bins. Again,
spectra for all three EPIC instruments were simultaneously fitted, using an
absorbed MEKAL model. Fits with hydrogen column fixed at the Galactic value
(5.00$\times$10$^{20}$ cm$^2$), and with column free to vary were
performed. The resulting profiles are shown in Figure~\ref{fig-Tprof}.

The first and most important feature of these profiles is that the group
appears to be isothermal. The temperature is consistent with that derived
from \asca\ \citep{Finoguenovetal01}. The abundance shows some evidence of
decline with increasing radius, but is consistent with the \asca\ Fe
abundance in all but the outermost bin. This is not unexpected, as the
\asca profile has only 2 bins in the range we are interested in, values
from which should be weighted averages of those we observe. The fits are
statistically acceptable in all bins. The low fit parameters in the three
outer bins reflect the reduced numbers of counts at larger radii. When the
Hydrogen column is free to fit it varies by a factor of $\sim$1.7, but does
not significantly improve the fits, except perhaps in the innermost bins.
The two innermost bins are also where \NH\ has the largest effect on the
fitted values of temperature and abundance, though the error regions are
consistent.

If the fitted \NH\ in the three innermost bins is taken at face value, the
excess hydrogen column above the measured Galactic value indicates a total
hydrogen mass of 1.64$\times$10$^{10}$\Msol\ within the cluster
core. Unfortunately this cannot be ruled out by radio measurements. Only an
upper limit on the neutral hydrogen mass is available from \textsc{Hi}
measurements, $\leq$6.8$\times$10$^9$\Msol, and strong radio emission from
NGC~6051 means that the error on this limit could be significant
\citep{ValentijnGiovanelli82}. However, if there is intrinsic absorption
then it must extend to a radius $>$55 kpc, approximately twice the \Dtf\
radius of the dominant galaxy. We cannot rule out an \textsc{Hi} cloud of
this size, but as it seems unlikely, and its inclusion does not greatly
improve our fits, we choose to assume that only Galactic absorption affects
the data.

Our central bin is relatively large, with a radius of 30\arcs\ ($\sim$18.5
kpc), and it is possible that we would not resolve a small cooling region
in the core of the dominant galaxy. NGC~6051 also hosts an AGN which, while
not apparent in the surface brightness profile, might be contribute to the
X-ray emission from the central bin. We therefore fitted additional models
to the spectra from this bin, including 2-temperature plasma models, plasma
+ power law models, and cooling models such as VMCFLOW and CEVMKL. These
were not successful. For the best fit two component models the
power law or high temperature plasma component was always poorly
constrained and made almost no contribution to the spectrum. The cooling
models suggest that the gas has a very narrow range of temperatures and is
certainly not cooling significantly. We therefore conclude that there is no
significant cooling or temperature structure in the cluster core, and that
the AGN in NGC~6051 must be heavily absorbed.

\begin{figure*}
\centerline{\epsfig{width=18cm,file=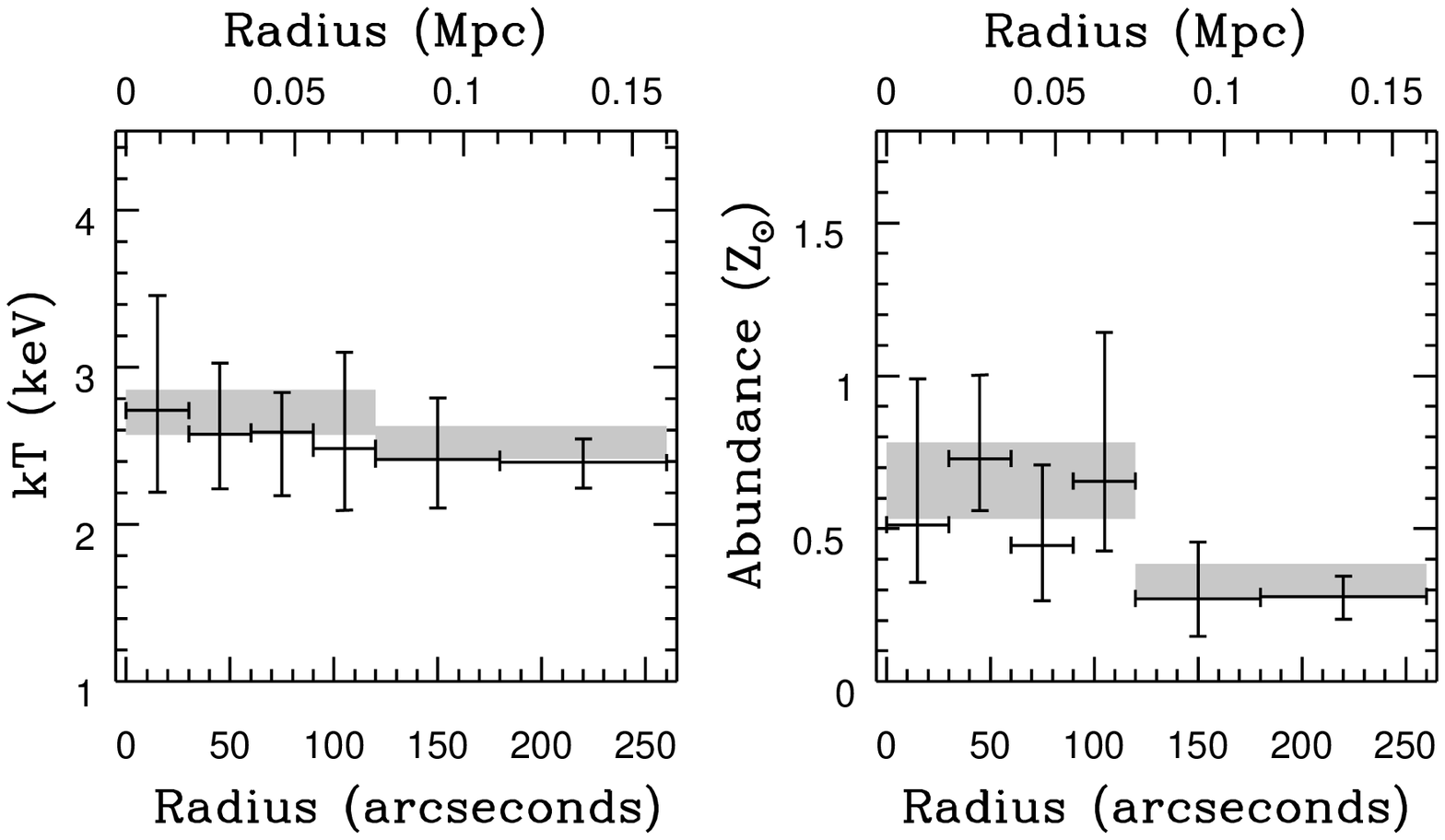,bbllx=20,bblly=480,bburx=592,bbury=779,clip=}}
\caption{
\label{fig-deproj}
Radial profiles of deprojected temperature and abundance. Crosses mark free
fits to the six bins, and shaded grey regions mark fit with the temperature
and abundance tied in the inner four and outer two bins. In both cases the
symbols show the 90\% error range on the fitted parameters.
}
\end{figure*}

Using the same annuli we also fitted a deprojected
spectral profile. The deprojection was carried out using the XSPEC
\textsc{projct} model in combination with an absorbed MEKAL model. The
hydrogen column was again held fixed at the Galactic value, and all metal
abundances were tied. The resulting deprojected temperature and abundance
profiles can be seen in Figure~\ref{fig-deproj}. As was to be expected, the
errors are quite large, and the same general trends are seen in the
projected and deprojected profiles. The temperature profile is consistent
with isothermality, while the abundance profile declines with radius. The
fits in the two outermost bins are somewhat better constrained than the
inner four, so we performed further fits with the six bins grouped into two
sets. The results of these fits are also shown in Figure~\ref{fig-deproj}
and the improvement in accuracy is clear. It is again clear that while
temperature is consistent with isothermality, showing only a slight trend
with radius, there is a decline in abundance at large radii.

\begin{figure}[!ht]
\centerline{\epsfig{width=9cm,file=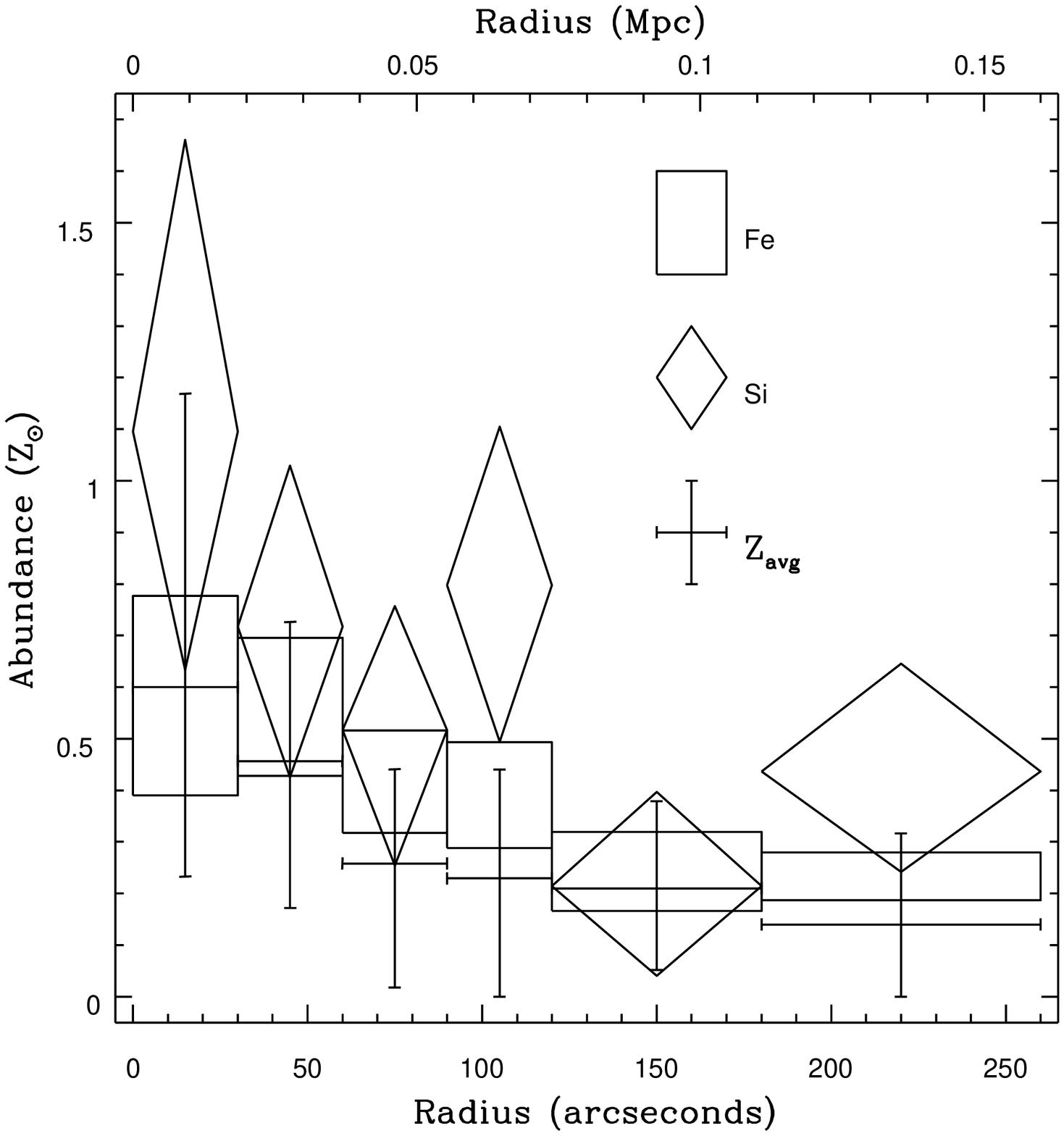,bbllx=20,bblly=200,bburx=592,bbury=779,clip=}}
\caption{
\label{fig-metals}
Radial abundance profiles for Fe, Si and all other metals combined. 90\%
errors on the fitted parameters are shown. \NH\ was frozen at the Galactic value.
}
\end{figure}

We also fit projected radial spectra with the abundance of Fe and Si free
to vary.  All other metals were tied, but free to vary as a group. Hydrogen
column was held fixed at the Galactic value. Given the lower numbers of
counts at large radii, it was necessary to merge the two outermost bins to
achieve a reliable fit. The results are shown in Figure~\ref{fig-metals}.
There is a marginal trend for a higher Si abundance in five of the six
bins, but the difference is only significant (at the 90\% level) in bin 4.
Overall we cannot claim any significant difference in abundance between Fe,
Si and other elements, and there is no trend with radius beyond the decline
already noted. The Fe and Si abundances are consistent with the \asca\ 
measurements of \citet{Finoguenovetal01} in the core, but are a poorer
match at larger radii. The outer bins are quite comparable to the
abundances derived by \citet{Fukazawaetal98}, also from \asca\ though these
were measured at slightly larger radii than our data permit. Given the
difficulties associated with \asca\ analysis, particularly in a field
containing a bright AGN, we consider the agreement acceptable.

\subsubsection{Spectral maps}

In order to look for correlations between the temperature structure of the
halo of AWM~4 and the radio lobes of NGC~6051, we prepared adaptively
binned temperature and abundance maps of the cluster. Appropriately cleaned datasets for
all three cameras, with point sources removed, were used
to create the source spectra, and the background in each case was taken
from similarly cleaned \citet{ReadPonman03} datasets. It should be noted
that in this case we did not use the ``double subtraction'' technique
described in Section~\ref{sec-obs}. This may lead to a bias toward lower
temperatures in the temperature map. However, as the map uses spectra with
relatively small numbers of counts compared to our large area spectra, this
bias is likely to be small compared to the statistical errors. 

As construction of a spectral map requires the extraction and fitting of a
large number of spectra, we used the \sas\ task \textsc{evigweight} to
correct the observation and background datasets. This task calculates a
statistical weight for each event in a dataset, effectively correcting the
data to account for the telescope effective area, detector quantum
efficiency and filter transmission. The corrected data therefore behave as
if the EPIC instruments were ``flat'', perfect detectors. This removes the
need for the calculation of Ancillary Response Files (ARFs) for every
spectrum, replacing them with a single on-axis ARF. Redistribution Matrix
Files (RMFs), which contain information which allows the true energy of
incident photons to be calculated, are still required. We used the
pregenerated ``canned'' RMFs available from the \xmm\ \sas\ calibration
webpage\footnotemark.

\footnotetext{http://xmm.vilspa.esa.es/external/xmm\_sw\_cal/calib/index.shtml}

\begin{figure*}
\centerline{\epsfig{file=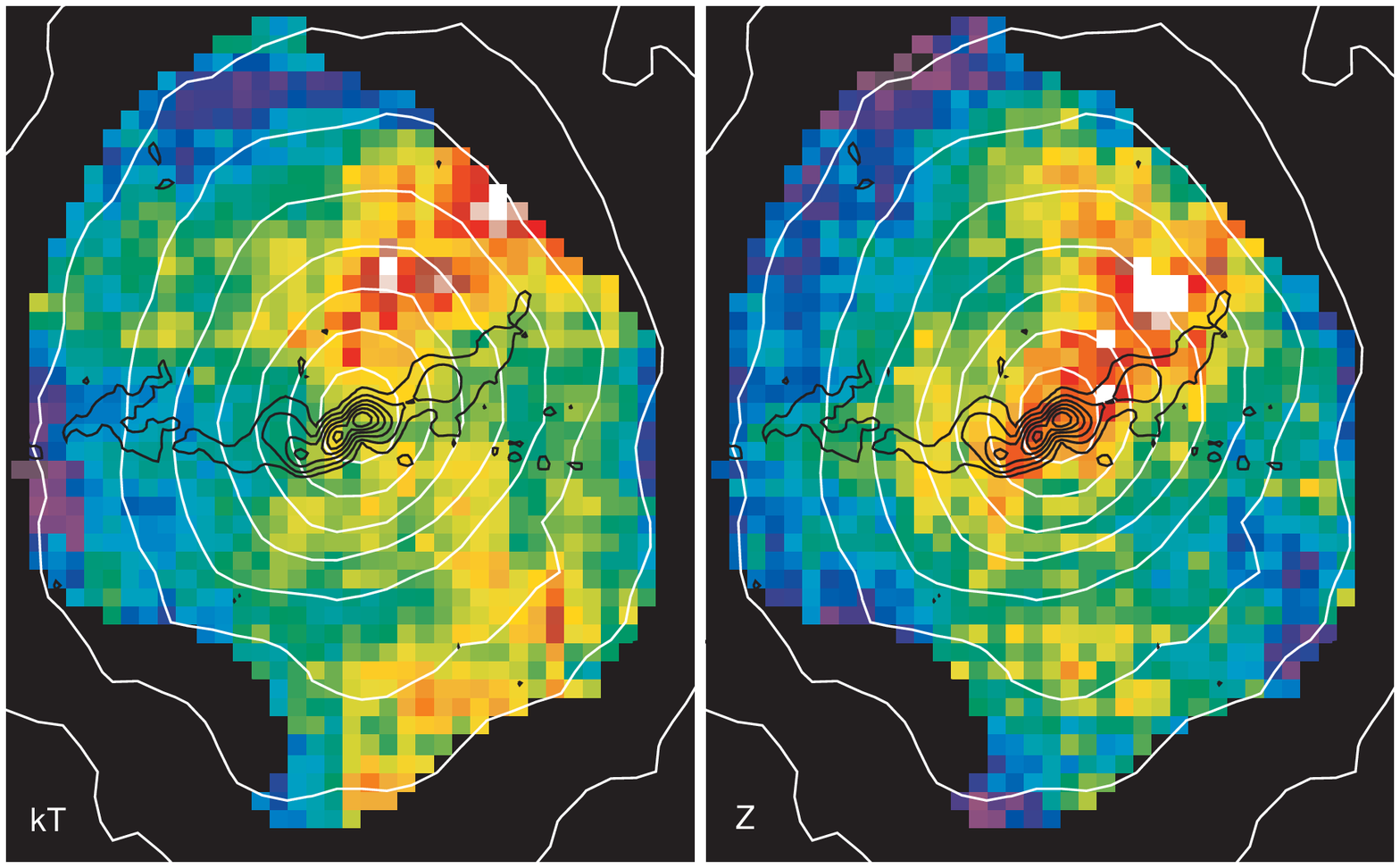,width=18cm}}
\centerline{\epsfig{file=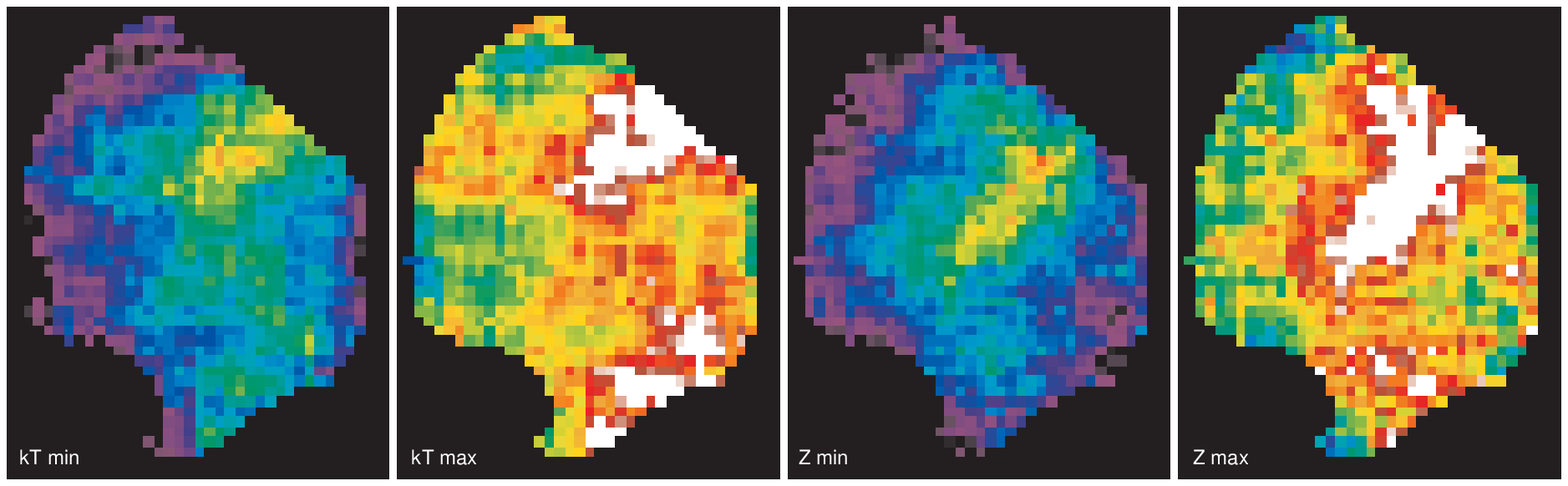,width=18cm}}
\caption{
\label{fig-Tmap}
Adaptively binned temperature and abundance maps of AWM~4, based on spectra
from all three EPIC instruments. The two upper panels show the best fit
values, while the lower panels show the 90\% error bounds. Each pixel
represents and adaptively chosen square region containing a total of 1600
counts, summed over the three cameras. Pixels for which the (90\%) error on
the temperature is larger than 20\% of the best fit value have been masked
out of the images. White contours show X-ray surface brightness, black
contours mark VLA-First 20 cm surface brightness. The maps are aligned so
that north is toward the top of the page and east is to the left. The pixel
size is 6.4\arcs. The colour scale in all images is linear. In the
temperature maps purple is $\sim$2.2 keV, blue 2.3-2.5, green 2.6-2.7,
yellow 2.7-2.8, red 2.9-3.1 and white 3.2 keV and higher. In the abundance maps
blue is 0.2-0.4 \Zsol, green is 0.5-0.6, red is 0.75-0.95 and white is solar abundance.}
\end{figure*}

The spectral regions for the map were selected using the following method.
A square region with side length $\sim$410\arcs\ was defined, centred on
the peak of the X--ray emission. This region was divided into a
64$\times$64 grid of pixels, each 6.4\arcs$^2$. Each pixel had an
associated spectral extraction region, with side length free to vary
between 19\arcs\ and 60\arcs. Note that this means that pixels in the map
correspond to regions centred on, but larger than, the pixels themselves,
and that these regions vary in size with the surface brightness of the
cluster.  The region size was optimised in each case to be as small as
possible while including at least 1600 source counts, summed over all three
EPIC cameras. Any pixel for which a 1200\arcs$^2$ region did not contain
1600 counts was ignored. For all pixels with at least 1600 counts, spectra
were extracted for source and background in all three cameras. The spectra
were grouped to have at least 30 counts per bin, and simultaneously fitted
with an absorbed MEKAL model. The creation of source and background
spectra, grouping and fitting were performed in \textsc{isis} v1.1.6
\citep{HouckDenicola00}.  Hydrogen column was held fixed at the Galactic
value (5$\times$10$^{20}$ cm$^{-2}$) and energies below 0.5 keV and above
8.0 keV were ignored. 90\% errors on all fits were calculated, and any
pixel with errors greater than 20\% of the best fit temperature was not
included in the map. The best fit value (or 90\% bound value) from each fit
is represented by the colour of the pixel in the final map, which can be
seen in Figure~\ref{fig-Tmap}.  Blue pixels represent low values, red
pixels high.

The map shows significant variation in temperature and abundance over the
central part of the cluster. Probably the most obvious features are the
regions of high temperature and high abundance extending to the northwest
of the cluster center. These are not perfectly cospatial, with the high
temperature gas lying slightly to the north of the high abundance gas. The
high abundance region extends into the core of the cluster, with high
abundance gas coincident with the central galaxy. The high temperature
region does not extend into the core. There is also some moderate abundance
gas to the east (left) of the core, which corresponds to a region of quite
low temperatures. The radio contours marked on Figure~\ref{fig-Tmap} appear
to correlate with some of the temperature structure. The eastern lobe lies
in a region of low temperature gas. The western lobe extends through a
region in which gas temperature decreases, and its northern edge
corresponds to the southern limit of the high abundance, high temperature
region. This may indicate that the radio lobes are interacting with the
X-ray gas, a possibility which we shall explore further in a later section.

As the spectral maps use a somewhat different fitting technique to our
other spectral fits, we carried out cross-checking to ensure the two
methods agree. Using the extraction regions for a small number of map
pixels, we extracted spectra and created ARFs, RMFs and ``double
background'' spectra. We then fit these in \textsc{xspec} and compared the
results to those found by the map fitting software. In all cases, the best
fit parameters were the same to within the (admittedly large) margin of
error. Using background spectra extracted from the \citet{ReadPonman03}
datasets with no correction for excess soft emission brought the fits into
closer agreement, but did not significantly alter the result in each pixel.
The maps only cover the brightest portion of the cluster core, so the
contribution from the soft background component is not expected to be
important.

\begin{figure}
\centerline{\epsfig{file=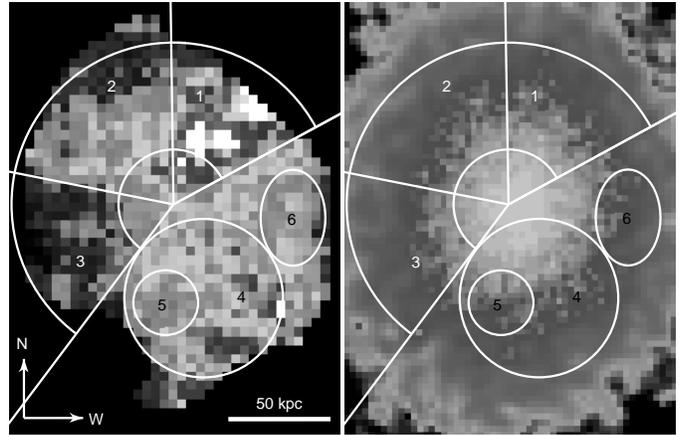, width=9cm}}
\caption{\label{fig-map_comp}
  \textit{left}: Temperature map of the core of AWM~4, as shown in
  Figure~\ref{fig-Tmap}. \textit{right}: Adaptively smoothed mosaiced image
  of AWM~4, using data from both MOS and PN cameras. Both images show the
  regions used to extract spectra for comparison between the map spectral
  fits and those used for the integrated spectra and spectral profiles.
  Region 4 is a large circle with a smaller circle, region 5, excluded.
  Further details are given in the text.  }
\end{figure}

As a further test, we selected regions in the spectral maps which appeared
to show particular features and then extracted spectra for those regions as
we would for our other spectral fits (generating ARFs and RMFs for each
region, using ``double background'' spectra). All spectra were fitted using
an absorbed MEKAL model with the hydrogen column fixed at the galactic
value (5$\times$10$^{20}$ cm$^2$). Spectra were grouped to 20 counts per
bin and fit in the energy range 0.5-8.0 keV. Figure~\ref{fig-map_comp}
shows the regions marked on the temperature map and on an adaptively
smoothed image of the cluster. We compared the fits to
these spectra with the range of temperatures and abundances found in the
spectral maps. Table~\ref{tab-map_comp} shows the results of the fits.

%%%%FIG MOVED FROM RADIAL PROFS ABOVE

%%%%%

\label{sec-maps}

\begin{table*}
\begin{center}
\begin{tabular}{lcc|cc|ccc|ccc}
 & \multicolumn{2}{c|}{Radii} & \multicolumn{2}{c|}{Map} & \multicolumn{3}{c|}{\textsc{sas v5.4.1}} & \multicolumn{3}{c}{\textsc{sas v6.0.0}} \\
Region & Inner & Outer & kT & Z & kT & Z & red. $\chi^2$/d.o.f. & kT & Z & red. $\chi^2$/d.o.f. \\ %& counts 
 & (\arcs) & (\arcs) & (keV) & (\Zsol) & (keV) & (\Zsol) & & (keV) & (\Zsol) & \\
\hline\\[-3mm] %& (bg sub.) 
1 & 0 & 120 & 2.7-3.2 & 0.35-1.0 & 2.91$^{+0.15}_{-0.14}$ & $0.69^{+0.13}_{-0.12}$ &
0.943/283 & 2.97$\pm$0.16 & 0.78$^{+0.17}_{-0.14}$ & 0.998/269 \\[1mm] %& 7292 

2 & 0 & 120 & 2.4-3.0 & 0.3-0.66 & 2.67$\pm$0.12 & 0.48$^{+0.09}_{-0.08}$ & 1.088/325 
& 2.69$\pm$0.13 & 0.44$^{+0.10}_{-0.09}$ & 1.130/305 \\[1mm] % & 8750 

3 & 40 & 120 & 2.3-2.7 & 0.3-0.75 & 2.46$^{+0.21}_{-0.18}$ & 0.44$^{+0.13}_{-0.10}$ & 1.048/207 & 2.64$^{+0.16}_{-0.15}$ & 0.51$^{+0.16}_{-0.11}$ & 0.984/212 \\[1mm] %& 5557 

4 & 14 & 130 & 2.6-3.1 & 0.3-0.8 & 2.82$^{+0.13}_{-0.12}$ & 0.50$^{+0.10}_{-0.09}$ &
1.048/328 & 2.82$^{+0.14}_{-0.13}$ & 0.50$^{+0.10}_{-0.11}$ & 1.062/309
\\[1mm] %& 9580 

5 & 48 & 95 & 2.55-2.75 & 0.5-0.7 & 2.74$^{+0.58}_{-0.41}$ & 0.80$^{+0.54}_{-0.33}$ &
1.031/61 & 2.78$^{+0.46}_{-0.33}$ & 0.57$^{+0.45}_{-0.23}$ & 0.857/63
\\[1mm] %& 1609 

6 & 63 & 110 & 2.4-2.8 & 0.4-0.7 & 2.87$^{+0.55}_{-0.43}$ & 0.55$^{+0.36}_{-0.24}$ & 0.969/71 & 2.74$^{+0.36}_{-0.33}$ & 0.41$^{+0.29}_{-0.20}$ & 0.901/71
\\[1mm] %& 1761 
\end{tabular}
\end{center}
\caption{\label{tab-map_comp}
Best fit parameters for fits to spectra extracted from the regions shown in
Figure~\ref{fig-map_comp}, using \textsc{sas v5.4.1} and \textsc{sas
  v6.0.0}. All fits were performed with hydrogen column fixed at the
Galactic value, and counts grouped into bins with a minimum of 20
counts. Only energies between 0.5 and 8.0 keV were used in the fits. The
maximum and minimum radius of each region from the surface brightness peak
are given in columns 2 and 3. Columns 4 and 5 show the approximate range of
values seen in the temperature and abundance maps. Columns 6, 7 and 8 show
results from fits to spectra extracted using \textsc{sas v5.4.1}
calibration, columns 9, 10 and 11 show results produced using \textsc{sas
  v6.0.0}. 90 per cent errors are shown for fitted parameters. 
}
\end{table*}

In regions 1-4, the fits are well constrained and agree well with the
values found in the map. Regions 5 and 6 are less well constrained, and
although the best fit temperature and abundance agree with those found in
the map to within the margin of error, the agreement is not particularly
good. Region 5 has a temperature at the high end of the range found in the
map, and a best fit abundance greater than the value found in the
map. Region 6 has a similar problem with temperature. However, both of
these regions have relatively low numbers of counts, around 500 in each of
the MOS cameras. They are also physically small, not much larger than the
PSF, and region 5 lies on a PN chip gap. We therefore do not believe that
these discrepancies are serious, and consider the spectral maps to be
acceptably accurate and reliable.

As stated in Section~\ref{sec-intro}, we performed our analysis using
\textsc{sas v5.4.1}. However, \textsc{sas v6.0.0} has since been released.
A major calibration improvement was incorporated in this release,
correcting the telescope vignetting functions of the EPIC instruments.
\citet{Lumbetal03} demonstrated that the optical axes of the telescopes
were not perfectly aligned with their boresights, as had been assumed,
meaning that objects apparently on axis were in fact slightly removed from
the true axis of the telescopes. As it seems possible that our spectral
maps could be altered by this improvement in calibration, we reprocessed
the data for AWM~4 using \textsc{sas v6.0.0}, and regenerated the spectral
maps. Comparison showed that the improved vignetting correction made little
or no difference to the final maps, with the various features appearing
unchanged.  As a further test, we extracted spectra and responses for the
six regions described in the previous paragraph. The results of fits to
these spectra are given in Table~\ref{tab-map_comp}. The fits are all in
reasonable agreement with those found for the \textsc{sas v5} spectra, and
with the spectral map. Region 5, which contains the smallest number of
counts, has a slightly higher temperature than the map (though with large
errors), but the abundance is in better agreement with the map than was the
case with the \textsc{sas v5} spectra. We therefore conclude that while the
changes to the vignetting functions may well produce minor changes in the
spectral fits to each map pixel, the features observed in the map are
likely to be real rather than products of poor calibration.

\subsection{Three-dimensional properties}
\label{sec-3d}
Given the well defined surface brightness profile and near isothermal
temperature profile of the cluster, we are able to use our models to
estimate some of the 3-d properties of AWM~4. Even the temperature map only
shows a 15 per cent variation between the hottest (region 1) and coolest
(region 3) areas. We use software provided by S.~Helsdon, which uses our
surface brightness and temperature models to infer the gas density profile.
We use the 2-component surface brightness model shown in
Table~\ref{tab-SB}, and assume an isothermal temperature profile with
$T$=2.6 keV. This temperature was determined by fitting a constant model
to our deprojected temperature profile. Using the projected profile would
not affect the choice of $T$ significantly. The density profile is
normalised to reproduce the X--ray luminosity of the cluster, determined
from our best fitting MEKAL model of the cluster halo. Given this density
profile, we can use the well known equation for hydrostatic equilibrium,

\begin{equation}
M_{tot}(<r) = -\frac{kTr}{\mu m_pG}\left(\frac{d{\rm ln}\rho_{gas}}{d{\rm
      ln}r}+\frac{d{\rm ln}T}{d{\rm ln}r}\right),
\end{equation}

\noindent to calculate the total mass within a given radius. From the gas
density and total mass, we can calculate parameters such as gas fraction,
cooling time and entropy, where entropy is defined to be

\begin{equation}
S = \frac{T}{n_e^{\frac{2}{3}}}.
\end{equation}

\begin{figure*}
\centerline{\epsfig{width=18cm,file=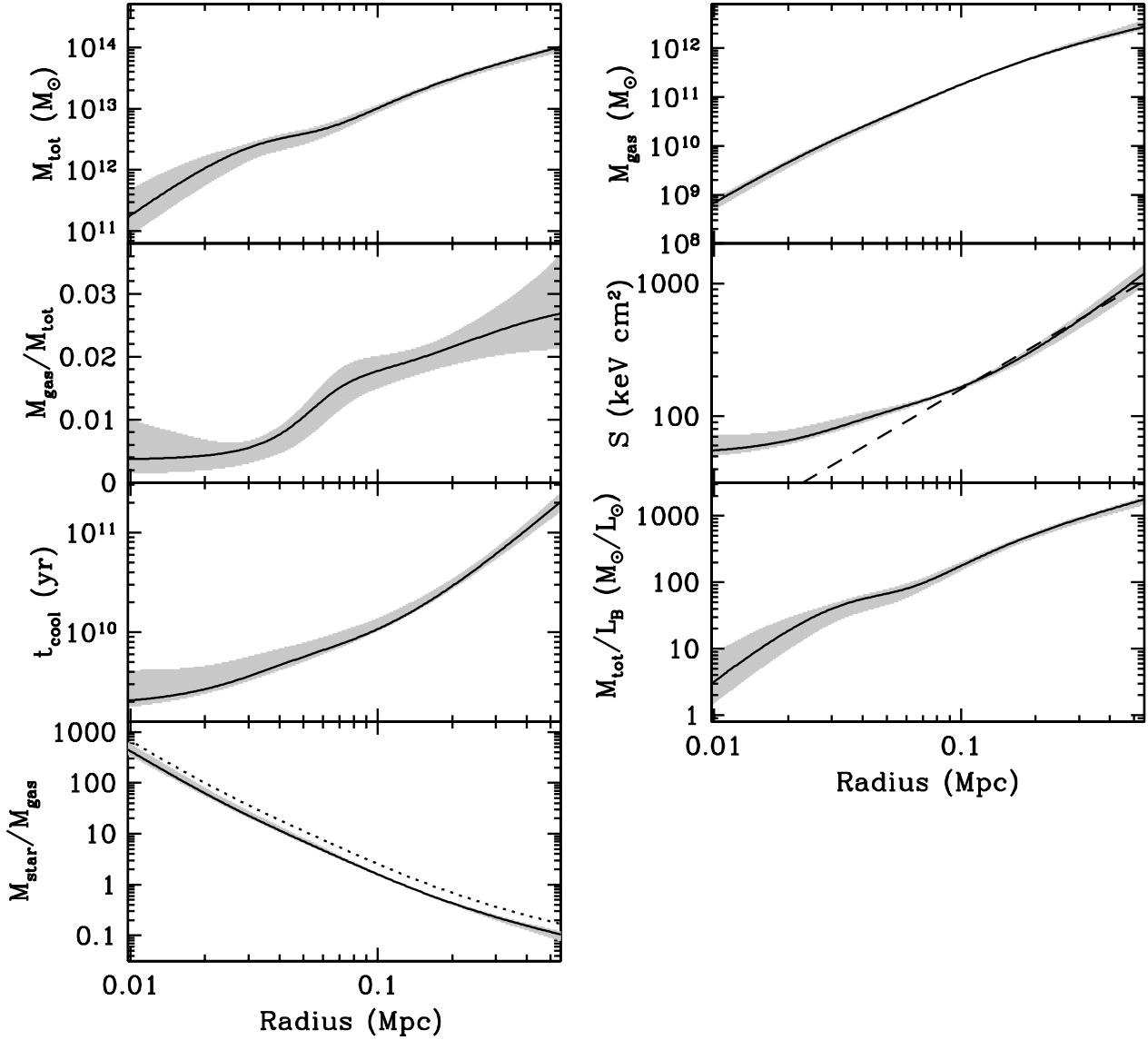,bbllx=20,bblly=230,bburx=592,bbury=739,clip=}}
\caption{
\label{fig-extrap} Total mass, gas mass, gas fraction, entropy, cooling
time, mass-to-light ratio and stellar mass to gas mass ratio for AWM~4.
1$\sigma$ errors are marked by the grey regions. The profiles are based on
data measured to $\sim$160 kpc, but we have extrapolated our models out to
0.5 Mpc. The dashed line in the plot of entropy shows the $S\propto
r^{1.1}$ behaviour expected from shock heating.  Stellar mass and optical
light contributions in the last two plots are from NGC~6051 only. In the
stellar mass to gas mass ratio plot, we assume that the mass-to-light ratio
of stars in NGC~6051 is either 5 $M_\odot/L_\odot$ (solid line) or 8
$M_\odot/L_\odot$ (dotted line). Note that we only include the contribution
of NGC~6051 to optical luminosity, so the true mass-to-light ratio of the
cluster probably never exceeds 200 $M_\odot/L_\odot$. Errors are plotted
for a stellar mass-to-light ratio of 5 only, so as to avoid confusion.  }
\end{figure*}

\noindent Errors on the parameters are estimated based on the measured
errors in temperature, surface brightness, etc, using a montecarlo
technique. We generate 10000 realisations of the profiles and then
determine the 1$\sigma$ upper and lower bounds for each parameter at any
given radius. Figure~\ref{fig-extrap} shows some of the parameters calculated.

The total mass profile indicates a slightly higher mass than that
calculated from the velocity dispersion of the galaxy population
\citep{KoranyiGeller02}, but is well within the errors on this optical
estimate. Previous mass estimates based on X-ray observations also agree
fairly well with our findings.  \citet{JonesForman99} and
\citet{Whiteetal97} estimate the mass based on \einstein\ data. The former
find masses of $\sim$10$^{14}$ h$^{-1}$\Msol\ within 0.5 Mpc and
2.6$\times$10$^{14}$ h$^{-1}$\Msol\ within 1 Mpc, while the latter find a
mass of 1.24$\times$10$^{14}$ h$^{-1}$\Msol\ within 1 Mpc. Using \rosat\ 
data, \citet{DellAntonioetal95} find a mass of 5.87$\times$10$^{13}$
h$^{-1}$\Msol\ within 0.42 Mpc. The estimates are quite consistent with
each other, at least to within a factor of 2, and also match our
gravitating mass profile fairly well. We find a mass of
$\sim$9$\times$10$^{13}$ h$^{-1}$\Msol\ within 0.5 Mpc.
The gas mass also appears to be quite comparable to other estimates.
\citet{JonesForman99} find $M_{gas}$ = (1.95$\pm$0.18)$\times$10$^{12}$
h$^{-2.5}$\Msol\ within 0.5 Mpc and (6.71$\pm$0.69)$\times$10$^{12}$
h$^{-2.5}$\Msol\ within 1 Mpc, \citet{Whiteetal97} estimate
$M_{gas}$ = 6.5$\times$10$^{12}$ h$^{-2.5}$\Msol\ within 1 Mpc and
\citet{DellAntonioetal95} find 3.6$\times$10$^{12}$ h$^{-2.5}$\Msol\ within
0.42 Mpc. We find $M_{gas}\simeq $2.5$\times$10$^{12}$ h$^{-2.5}$\Msol\ 
within 0.5 Mpc.

If we assume a virial radius of $\sim$1.2 Mpc (see Section~\ref{sec-AGN}
for discussion of this choice), the gas fraction at 0.1$\times R_{virial}$
is $\sim$2 per cent, very similar to that found for other systems of
similar temperature \citep{Sandersonetal03}. The gas fraction at larger
radii is somewhat low, but given the extrapolation required to reach the
virial radius, both for our data and that used in other studies, some
degree of error must be expected. Entropy at 0.1$\times R_{virial}$ is also
quite comparable to that found in other systems \citep{Ponmanetal03}.
Following \citep{PrattArnaud03}, we have plotted the $S\propto r^{1.1}$ law
predicted by numerical modelling of shock heating in a spherical collapse
\citep{TozziNorman01} and normalized to the entropy of a 10 keV cluster
from \citet{Ponmanetal03}. It is clear that the entropy profile is in close
agreement with this prediction, with the profile flattening only within
$\sim$0.1 Mpc ($\sim$0.08$\times R_{virial}$). This is very similar to the
behaviour reported for A1983 and A1413 by Pratt \& Arnaud. However, the
cooling time is rather long, 2$\times$10$^9$ yr within NGC~6051, a factor
of 10 longer than in other relaxed clusters. This suggests that some
process has heated the gas or is maintaining its temperature, an issue we
will return to in Section~\ref{sec-discuss}.

\subsection{Metal masses and supernova enrichment}
Given the abundances measured in Section~\ref{sec-spectral} and the gas mass
profile calculated in Section~\ref{sec-3d}, we can estimate the masses of
different elements in the intra-cluster medium (ICM). We use our best fitting
APEC model of the cluster as a whole and the MEKAL fits to the six radial
bins shown in Figure~\ref{fig-metals}. The resulting masses are given in
Table~\ref{tab-metals}.

\begin{table}
\begin{center}
\begin{tabular}{l|cccc}
Spectral & \multicolumn{4}{c}{Mass (\Msol)} \\
Bin & Fe & Si & S & O \\
\hline
Total & 8.03$\times$10$^6$ & 9.54$\times$10$^6$ & 2.07$\times$10$^6$ & 7.38$\times$10$^7$ \\
1 & 1.02$\times$10$^5$ & 1.55$\times$10$^5$ & - & - \\
2 & 4.68$\times$10$^5$ & 4.56$\times$10$^5$ & - & - \\
3 & 6.28$\times$10$^5$ & 5.87$\times$10$^5$ & - & - \\    
4 & 8.56$\times$10$^5$ & 1.33$\times$10$^6$ & - & - \\
5 & 1.50$\times$10$^6$ & 1.03$\times$10$^6$ & - & - \\
6 & 2.75$\times$10$^6$ & 3.72$\times$10$^6$ & - & - \\
\end{tabular}
\end{center}
\caption{
\label{tab-metals}
Mass (in solar units) of Fe, Si, S and O for our best fitting spectral
model of the cluster, and for six radial bins. Bin 1 is in the cluster
centre, bin 6 is the outermost.
}
\end{table}

We can also use these masses to estimate the numbers of SNIa and SNII
needed to enrich the ICM. We assume yields of Si and Fe for type II
supernovae of y$_{Fe}$=0.07\Msol\ and y$_{Si}$=0.133\Msol\ 
\citep{Finoguenovetal00}, and from type Ia supernovae y$_{Fe}$=0.744\Msol\ 
and y$_{Si}$=0.158\Msol\ \citep{Thielemannetal93}. For the cluster as a
whole, out to a radius of 260 kpc, we estimate that $\sim$42\% of the Fe
and $\sim$7\% of the Si in the halo was injected by type Ia supernovae. In
general the spectral profile suggests that SNIa dominate Fe production in
the inner part of the cluster, with $\sim$55-57\% of the Fe in bins 2 and 3
being produced by SNIa while 20-30\% of the Fe in bins 5 and 6 is produced
by this type of supernova. However, the high Si value found for the central
bin reverses this trend, suggesting that only $\sim$22\% of Fe in this bin
was produced in SNIa. We note that the large errors on the Si abundance in
this bin make the result somewhat uncertain, however, and that the
innermost spectrum is poorly fit with a single temperature model, making
the derived abundances less certain. It is therefore likely that the
abundances in this bin are in fact compatible with a relatively high
fraction of type Ia supernovae.

It has been suggested that more complex models of supernova enrichment,
involving two (or more) types of SNIa and revised elemental yields for
SNII \citep[e.g.,][]{Finoguenovetal02a}, can produce a more accurate
picture of the enrichment history of groups and clusters. In AWM~4, it is
notable that the total mass of O and S for the system do not fit into the
simple scheme of enrichment presented above, and would suggest a different
ratio of supernovae types if we were to use them instead of Si. However,
attempts to fit all four elements simultaneously using models including
multiple types of SNIa were not successful, primarily because of the large
mass of O. This may indicate that the masses derived from the integrated
spectrum are not accurate, which would not be surprising given the
variation in abundance with radius. We believe that the abundances measured
in radial bins are a more accurate measure of the true abundance, but
unfortunately the spectra from these bins do not support measurement of
more elements, restricting us to the simple supernova models discussed above.

\section{Discussion}
\label{sec-discuss}

\subsection{Temperature and Abundance structure}
\label{sec-TZstruct}
The fits presented in Section~\ref{sec-spectral} show that the inner
$\sim$150 kpc of AWM~4 are relatively isothermal, with a mean temperature
of $\sim$2.5 keV. There is a possible slight rise in temperature toward
the core of the group, but the trend for increasing abundance toward the
center is much stronger. Metallicity rises from 0.3 \Zsol\ at 150 kpc to
$\sim$0.7 \Zsol\ in the inner parts of NGC~6051. However, these fits use
spectra integrated over relatively large areas. The spectral maps shown in
Section~\ref{sec-maps} use spectra extracted from rather smaller regions,
and show considerably more structure than the azimuthally averaged radial
spectral profiles. The errors bounds on the values in the maps, and the
fits we carried out to check the accuracy of the maps, suggest that much of
this structure is real and significant. This does not indicate that the radial
profiles are incorrect, but shows the effect of azimuthal averaging on the
spectra. The spectral maps correspond to the four inner bins of the radial
profiles, and the error bounds on the values in each bin are quite
comparable to the range of temperatures and abundances seen in the maps. It
is clear though that without the maps we would miss some important
structure in the cluster. 

The regions shown in Figure~\ref{fig-map_comp} were chosen to include some
of the features which are visible in the spectral maps. These can be
describes as follows. Region 1 contains the high temperature and high
metallicity emission which extends to the northwest of the cluster
core. Both temperature and abundance show ``ridges'' of high values, but
the peak of the two ridges does not coincide; the highest abundance region
runs from the core of NGC~6051 to the northwest, while the peak temperature
runs parallel to it but slightly further north, and does not enter the
galaxy core. Region 3 is somewhat cooler than most of the rest of the map,
and lies to the east of the core, along the line of the eastern radio
jet. Region 2 lies to the north of the core, between regions 1 and 3, and
has fairly average temperature and abundance. Region 4 covers the area south
of the core, and contains some areas of high temperature and moderate
abundance. In the eastern part of region 4, we have separated out a smaller
region (5) where the temperature is quite low. Region 6 is an area of
moderate to low temperature and abundance to the west of the core. 

Regions 5 and 6 both have considerably poorer statistics than the other
four regions, as they are smaller and contain fewer counts. Region 5 in
fact lies over a chip gap in the PN camera, which reduces the number of
available counts considerably. This lack of counts in the PN may be at
least partially responsible for the cool feature which region 5 was chosen
to highlight, and the rather large errors suggest that the feature should
be treated with some caution. However, region 6 does not contain chip gaps
or regions removed because of point source emission and although the
temperature errors for the fit to this region are comparable to those for
region 5, the abundance error is slightly smaller. We therefore consider
region 6 to be reliably fitted. 

\subsubsection{Interaction of the radio jets with the ICM}
The three main features of the map are the high temperature/abundance areas
in region 1, the low temperatures in region 3 and the moderate/high
abundances in region 4. Regions 1 and 2 appear to be correlated with the
position of the radio jets originating from NGC~6051. To the east, the
radio jets points toward and extends into the cool gas of region 3. To the
west, the jet extends along the southern boundary of region 1, coincident
with the edge of the ridge of high abundance and slightly to the south of
the highest temperatures. One explanation of these features are that they
are caused by the interaction of the radio jets with the X-ray emitting
plasma of the cluster halo.

\chandra\ observations of massive clusters which contain powerful radio
sources have shown that the interaction between jets and surrounding gas
can be powerful and complex. Perhaps the best cases are the Perseus cluster
\citep{Fabianetal02a} and M87 in the Virgo cluster \citep{Youngetal02}. In
Perseus, the radio emission from the radio jets corresponds with areas of
low X-ray emission in the inner halo. These are thought to be cavities in
the X-ray plasma hollowed out by the pressure of the radio-emitting plasma
from the AGN jets. The X-ray gas at the edges of the cavities is cool,
indicating that the expansion of the bubble of radio plasma inside has not
been rapid enough to cause strong shocks. However, very deep imaging shows
density waves in the ICM which are likely to be mild shocks and sound waves
caused by the creation of the cavities \citep{Fabianetal03}. In M87,
cavities are again visible to some extent, but X-ray emission is also
enhanced along the radio jets and at the hot spots in the radio
lobes. Numerous other features visible in the radio emission correspond to
bright areas in the X-ray, again suggesting that the radio plasma can
locally compress the ICM, raising its emissivity and surface brightness.

In AWM~4, the cavities seen in more massive clusters may provide an
explanation for the cool feature in region 3. The eastern radio jet may
have created a cavity in the X-ray halo. If so, the hotter gas which would
usually occupy this region would have been displaced, so that we see only
the slightly cooler gas in front of and behind it. We see no sign of such a
cavity in the X-ray images of the cluster, nor in the residual images which
result from subtracting our best fitting surface brightness models from
this image. However, the density and (line of sight) depth of the ICM in
AWM~4 is considerably smaller than in massive clusters such as Perseus and
Virgo, and so we might expect such a cavity to stand out less clearly.
AWM~4 is also more distant, the radio source in NGC~6051 is less powerful
and we are using a relatively short \xmm\ observation, with poorer spatial
resolution than is available from \chandra, all factors which would make
such a feature more difficult to detect. Another possibility is that region
3 is cooler because it contains cool gas drawn out from the galaxy core by
the action of the jets, as has been suggested for the regions around the
jets of M87 \citep{Bohringeretal01}, but the average to low abundance
measured in this region argues against this.

The western radio jet might be responsible for the high temperatures in
region 1, if it is capable of heating the gas through strong shocks. In
order to estimate the shock strength required, we use the standard
Rankine-Hugoniot jump conditions \citep[e.g.,][]{LandauLifshitz59} to
estimate the Mach number of the shock from the temperature difference
between shocked and unshocked gas. Regions 6 and 2 seem to be relatively
unaffected by the radio source, so either could be used to give the
temperature of the unshocked plasma. However, as well as being small,
region 6 is where we would expect to see a cavity if the radio source was
symmetrical, so we have chosen to use region 2. The temperature difference
between regions 1 and 2 (0.24 keV) could be produced by a shock of Mach
number $\mathcal{M}=1.09$. This would correspond to a density increase
across the shock front of a factor of $\sim$1.2, and a corresponding
increase in X-ray surface brightness of $\sim$1.4. As the cluster halo is
elliptical, with its major axis running roughly north-south, this change is
less obvious than it would be if there were a shock front extending due
west or east. We estimate that the surface brightness difference between
regions 1 and 2 owing to the shape of the halo is a factor of $\sim$1.15,
so the increase from any shock will be only marginally above this. We do in
fact see a difference between the two regions of a factor of 1.2, and the
error associated with the estimate is large enough that we cannot be
certain that the gas has been shock heated. It should also be noted that if
the shock is not moving in the plane of the sky (i.e. the radio jet does
not lie in this plane) then our estimate of the expected change in surface
brightness should be lowered. We calculate that an angle of only 15$^{\deg}$
would be sufficient to make the density jump indistinguishable from that
observed.

There are problems with the hypothesis of a shock. The most important is
that we do not see any evidence of such a shock to the east, in or around
region 3. As we have already mentioned, the interaction of jets with the
ICM of more massive clusters does not seem to be accompanied by shock
heating, so if this is occurring in AWM~4 it would be an interesting case
for further investigation. Similarly, shock heating would not provide an
obvious explanation of the high abundances seen along the northern edge of
the western jet. The northern jet of Centaurus A has been shown to
intersect an \textsc{Hi} cloud, compressing the gas and triggering star
formation \citep{Graham98}. A degree of star formation in the ICM of AWM~4
could in principle inject metals into the gas through supernovae and
stellar winds, but given the observed abundances (as high as solar) it
seems unlikely that sufficient numbers of stars could be formed without
being detected in the optical.

\subsubsection{Movement in the plane of the sky}
One further possibility which should be considered is that either NGC~6051
or other galaxies in the cluster have a significant velocity in the plane
of the sky. From the relative velocities of the galaxies in the line of
sight, the cluster appears to be relaxed, and the recession velocity of
NGC~6051 agrees well with the mean velocity of the cluster. However, the
cluster is quite extended to both north and south, so it is possible that
galaxies might have significant velocities which we cannot measure. If so,
a galaxy moving through the cluster core at relatively high speeds might
create a gaseous wake through either ram-pressure
\citep{GunnGott72} or viscous stripping \citep{Nulsen82}. A galaxy moving
on a near-radial orbit from south to north or northwest might produce
features which would pass through regions 4 and 1. We would expect
such a wake to have high abundance, as the gas would have been enriched by
stellar winds and supernovae in the galaxy. We would also expect the gas to
be relatively cool, as even in early-type galaxies with X-ray emitting
halos, temperatures greatly in excess of 1 keV are rare
\citep{Brownbreg98,Matsushita00}. This is a problem, as we observe
temperatures of 2.5 keV and above, in general hotter than the surrounding
medium. Alternatively, NGC~6051 could be moving, a possibility which is
supported by the angle between the radio jets. Movement of the galaxy to
the south or southwest might produce such an angle, as the pressure of the
ICM prevents the jet from moving at the same velocity as the galaxy. It is
more difficult to imagine what we would expect to observe were this the
case, but it seems possible that the hotter gas to the north of the jets
could have been shock heated by the jets before the galaxy moved south to
its current position. As a final option, we note that if the galaxy were
moving southeast instead of south, the high abundance ridge extending from
the core to the northwest might be explained as a tail of enriched gas much
like that seen in Abell~1795 \citep{Markevitchetal01}. In A1795 this gas
is cooler than its surroundings, but in AWM~4 it would have been affected
by the eastern radio jet, and possibly heated. Movement of NGC~6051 in
either direction does not explain the hot gas in region 4, and as we cannot
measure movements in the plane of the sky, these possibilities must remain speculative.

\subsection{Abundance - comparison with other systems}
A number of other groups and poor clusters have been observed with \xmm\ 
and have high quality abundance measurements available. The reflection
grating spectrometers (RGS) provide the greatest precision in
measurement, as individual line strengths can be modeled. However, the
EPIC cameras also provide excellent abundance estimates, and allow the
variation of metallicity with radius to be measured.

Systems for which high quality EPIC observations are available include
NGC~1399 \citep{Buote02} in Fornax, NGC~5044
\citep{Buoteetal03a,Buoteetal03b} and M87 in the Virgo cluster
\citep{Matsushitaetal03}. All three of these systems differ from AWM~4 in
that they show evidence of multi-phase gas in their cores and are best fit
by models with at least two temperature components. However, provided their
spectra are sufficiently well described by these models (and providing we
have accurately modeled our spectra) it is possible to compare the
abundances and their trend with radius. Our findings can be summarised as
follows; for the integrated spectrum we find that Si is the most abundant
element (of those we fit individually), followed by Fe, S and O. All are
subsolar, and all but Si have abundances $<$0.5\Zsol. Radial profiles
suggest that abundance increases toward the core of the group.

M87 shows a similar trend with radius, but considerably higher abundances.
Si, S and Fe peak at almost twice solar, with Si slightly more abundant
than Fe and S. We note that \citet{Matsushitaetal03} use the abundance
ratios of \citet{Feldman92}, which produce slightly higher Fe abundances
than those of \citet{AndersGrevesse79}, which we use. Were the authors to
use the Anders \& Grevesse abundances, Fe abundance would be reduced with
respect to Si, a situation much like that which we observe in AWM~4. O is
considerably less abundant, with a peak value of $\sim$0.7\Zsol. At radii
$<$1\arcm\ the profiles turn over and decrease in the galaxy core. This may
be caused by rapid cooling through line emission of gas in the high density
environment of the cluster core, which will preferentially remove gas with
high abundances. However, at the distance of M87, 1\arcm$\simeq$4.63 kpc,
too small a distance for us to resolve in AWM~4. Alternatively, it may be a
product of incorrect abundance estimates, caused by the multiphase nature
of the gas in the inner core.

The analysis of NGC~1399 shows it to have very supersolar Fe and Si, with
abundances increasing in the galaxy core. Si has a peak value of
$\sim$1.7\Zsol. Once again, Fe is measured using an abundance ratio which
differs from that which we use, but corrected to our standard the peak Fe
abundance is $\sim$1.5\Zsol. NGC~5044 is much like M87, with Fe and Si
showing solar or slightly supersolar peak abundances, and abundance rising
to a peak before falling in the core. S has a peak abundance of
$\sim$0.8\Zsol\ and does not drop in the inner bins, and O has a low
abundance ($\sim$0.2\Zsol) at all radii.

For Iron and Silicon, we can calculate the mass of metals found in each
system, allowing a direct comparison. For M87,abundance and gas mass
profiles are both based on \xmm\ analysis
\citep{Matsushitaetal02,Matsushitaetal03}, as are the abundances in the
other two systems \citep{Buote02,Buoteetal03b}. The gas mass profile for
NGC~1399 is taken from the \rosat\ HRI study of \citet{Paolilloetal02}, and for
NGC~5044 we use a profile based on \rosat\ PSPC analysis
\citep{Sandersonetal03}. All three galaxies have high quality abundance
measurements to at least 50 kpc radius, and this is the radius within which
we calculate metal masses. Table~\ref{tab-lpd} shows the masses of Fe and
Si.

\begin{table}
\begin{center}
\begin{tabular}{lcc}
System & Fe mass & Si Mass \\
 & (\Zsol) & (\Zsol) \\
NGC 1399 & 4.0$\times$10$^5$ & 3.4$\times$10$^5$ \\
NGC 5044 & 1.8$\times$10$^6$ & 1.5$\times$10$^6$ \\
M87 & 2.0$\times$10$^6$ & 2.5$\times$10$^6$ \\
AWM 4 & 1.2$\times$10$^6$ & 1.2$\times$10$^6$ \\
\end{tabular}
\end{center}
\caption{\label{tab-lpd}
Mass of Fe and Si within the central 50 kpc of each system.}
\end{table}

The most obvious differences between these results and ours is the high
abundances found in NGC~5044 and M87, and the low metal masses in NGC~1399.
The three galaxies have similar optical luminosities to NGC~6051,
suggesting that the stellar component in the core of each system is of
comparable mass. This suggests that enrichment of the ICM by supernovae
should be similar, yet we find differences of at least a factor of two. One
possible explanation could be that in the rich Virgo cluster numerous
smaller galaxies have passed through the cluster core and have been
stripped of their gas either by ram-pressure or tidal forces. This enriched
gas is then left in the surroundings of the dominant galaxy, where we
observe it. This seems less likely in the case of NGC~5044, however, as
this group is poorer than AWM~4; the group has 9 members \citep{Garcia93}
compared to 28 in AWM~4 \citep{KoranyiGeller02}. An alternative is the
model suggested by \citet{Bohringeretal04} in which, during periods of
relative stability, the cluster dominant galaxy builds up its own metal
rich ISM and a surrounding region of partially enriched mixed ISM/ICM gas.
In this model only cluster mergers disrupt this structure and thoroughly
mix the gas, so the degree of enrichment is dependent on the time since the
last major merger, and the differences we see between systems would reflect
differences in their merger histories. This can also be considered as the
age of the cool inner core, as during the periods between mergers gas will
radiatively cool and presumably be reheated. This model may well be
substantially correct, though our results from AWM~4 suggest that AGN
activity may also produce gas mixing (on a smaller scale than cluster
mergers), reducing the observed abundance peak in the cluster core.

Another alternative is that the multi-phase gas in the three systems
discussed above is a symptom of a cooling flow in which the gas is poorly
mixed. If the high abundance gas produced by supernovae and galaxy winds
does not mix well but instead remains in small clumps, then the high
abundance gas will cool much faster than the gas surrounding it.
\citet{MorrisFabian03} show that such a model can produce very large
abundances in the core, surrounding a small low abundance region in which
most of the high abundance gas has cooled out of the X-ray regime. In
AWM~4, the action of the AGN might have prevented such cooling by reheating
gas in the core, or by acting to mix the gas more thoroughly. However, we
note that this model is not supported by detailed modeling of EPIC and RGS
spectra \citep{Petersonetal03,Kaastraetal04}. A final possibility is that
the AGN in NGC~6051 is sufficiently effective in mixing the gas that even
without abundance inhomogeneities it has suppressed the abundance gradient
in the cluster core. Models of rising bubbles \citep[designed to study the
interaction of AGN and ICM around M87)][]{Churazovetal01} suggest that the
movement of bubbles out through the halo can entrain material and draw it
out from the core regions to larger radii. Entrained gas will likely fall
back into the core, as it is denser than the surrounding material, but this
might offer opportunities for mixing the gas and suppressing the abundance
gradient. The strongest argument against this possibility is the abundance
gradient in M87. The interaction between radio jets and ICM in that system
is clear, but the abundance gradient is still strong, arguing that the gas
does not mix. Clearly more detailed modeling of the processes of heating
and gas motions produced by AGN activity is required before these issues
can be explored further.

\subsection{Comparison with MKW 4}
\label{sec-AGN}

%%%%%Figure inserted from below%%%%%%%
\begin{figure*}
\centerline{\epsfig{width=18cm,file=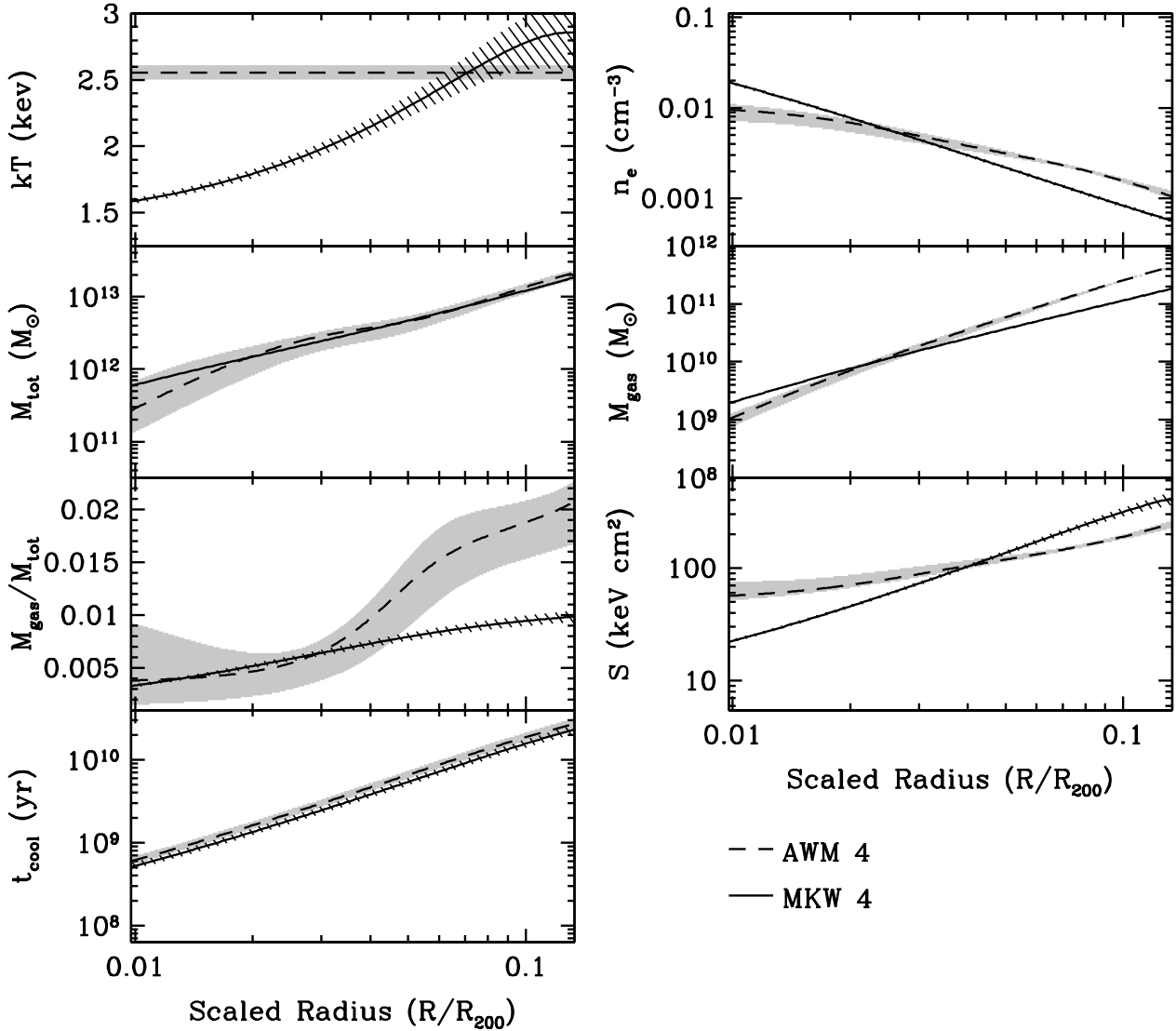,bbllx=20,bblly=250,bburx=592,bbury=739,clip=}}
\caption{
\label{fig-AWMvsMKW}Profiles of X-ray temperature, density, total mass, gas
mass, gas fraction, entropy and cooling time for AWM~4 and MKW~4, using
data from this paper and \citet{OSullivanetal03}. The profiles are
normalised to \Rth\ to correct for the differing masses and sizes of the
two clusters. AWM~4 is marked by a dashed line, with the 1$\sigma$ error
region marked in grey. MKW~4 is marked by a solid line, with the error
region hatched. On some plots the 1$\sigma$ errors are smaller than the
line width.  }
\end{figure*}
%%%%%%%%%%%%%%%%%%%%%%%%%%%%%%%%%%%%%%

AWM~4 contrasts markedly with MKW~4, a poor cluster of similar mass also
recently observed by \xmm\ \citep{OSullivanetal03} and \chandra\ 
\citep{Fukazawaetal04}. MKW~4 is relatively cool at large radii, with a
temperature of 1.6~keV at 250~kpc \citep{Helsdonponman00}. The temperature
rises to a peak at 100~kpc, then falls again toward the
core, to a central temperature of only $\sim$1.4~keV. Abundance within this
cooling region is considerably higher than at comparable radii in AWM~4, up
to $\sim$1.4\Zsol\ within the innermost 20~kpc. The abundance of Si and S
in this inner region is well in excess of that of Fe, demonstrating that
the gas in the inner core has been enriched by supernovae and has remained
undisturbed for some time. MKW~4 is not, however, well described by a
steady state cooling flow model. Spectral fitting using the EPIC
instruments suggests that there is little or no gas at low temperatures,
and the RGS data suggests a minimum temperature of $\sim$0.5 keV.

Although the general trend of the temperature profile in MKW~4 seems clear,
there is some discrepancy between results for the peak temperature. Our
\xmm\ analysis in a previous paper \citep{OSullivanetal03} suggests a peak
temperature of $\sim$3 keV. This is in reasonable agreement with results
from a \rosat\ observation \citep{Helsdonponman00}, which found the
temperature to be $\sim$2.4 keV at $\sim$45-80 kpc, and poorly defined but
above 3 keV at $\sim$80-125 kpc. However, results from \asca\ 
\citep{Finoguenovetal00} contradict these, and the analysis of \textit{XMM}
and \chandra\ data by \citet{Fukazawaetal04} finds a peak temperature of
only $\sim$2 keV. One possible explanation of for this is that the
background used in our analysis of the \textit{XMM} data for MKW~4 was
incorrect, causing over-subtraction at low energies and overestimation of
temperatures. It also seems possible that the opposite is true of the
Fukazawa et al. analysis. We have therefore decided to consider both
versions of the temperature profile in our comparison with AWM~4. The model
fit to our \textit{XMM} temperature profile is described in
\citet{OSullivanetal03}. We fitted the combined \chandra\ and \xmm\ 
temperature data from Fukazawa et al. with a 'Universal' temperature
profile of the form described by \citet{Allenetal01b}, though not
normalized to the R$_{2500}$ overdensity radius. This fit has an outer
temperature of 1.76$^{+0.07}_{-0.06}$ keV, core temperature of
1.35$^{+0.03}_{-0.05}$ keV, core radius of 8.68$^{+0.88}_{-1.72}$ kpc, and
slope parameter $\eta$=4.63$^{+3.28}_{-2.23}$. We note that model produces
a flat temperature profile with kT=1.76 keV outside $\sim$35 kpc, and so
can be considered a very conservative alternative to our \textit{XMM}-only
model. 

%%%%%Figure moved up%%%%%%%

Figure~\ref{fig-AWMvsMKW} shows a comparison of some of the 3-dimensional
model properties of the two clusters, based on our \textit{XMM} analysis.
We have scaled the models using the \Rth\ radius of the clusters (the
radius within which the average density of the cluster is 200 times the
critical density of the universe, equivalent to the virial radius). For
MKW~4, we use \Rth=0.97 \hs$^{-1}$ Mpc \citep[see][for a discussion of this
value]{OSullivanetal03}.  \citet{Sandersonetal03} use \rosat\ and \asca\ 
data to estimate \Rth\ for AWM~4, and find \Rth=1.44$^{+0.32}_{-0.27}$
\hs$^{-1}$ Mpc. Extrapolating from our surface brightness models and
assuming a constant temperature, we estimate \Rth=1.18 \hs$^{-1}$ Mpc,
similar to the value we would obtain based on the work of
\citet{NavarroFW95}, who found \Rth=1.17 \hs$^{-1}$ Mpc. There are
potential problems with using any of these estimates.  Estimates based on
an isothermal temperature profile will not account for any temperature
gradient at larger radii, while the \asca\ and \rosat\ analysis assumed a
central cooling region and therefore may not model the inner halo
accurately. However, the estimates are in broad agreement, and we choose to
use the estimate based on the \xmm\ data, noting the inherent uncertainty.

\begin{figure}
\centerline{\epsfig{file=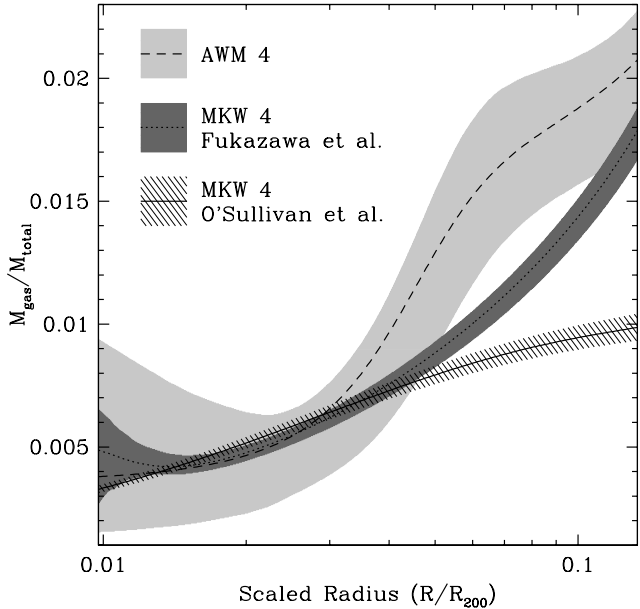,width=8.5cm,bbllx=20,bblly=210,bburx=560,bbury=739,clip=}}
\caption{\label{fig-Gfrac}Gas fraction profiles for AWM~4 and MKW~4. The
  MKW~4 profiles were determined using model fits to either the
  \citet{Fukazawaetal04} \textit{XMM}+\chandra\ temperature profile or the
  \citet{OSullivanetal03} \textit{XMM}-only profile, which has a higher
  peak temperature.}
\end{figure}

A second comparison, using the Fukazawa et al. temperature profile,
produced very similar results. As expected, gas density and gas mass, which
are dependent primarily on the surface brightness profile, were
essentially unchanged, as was the cooling time. The total mass profile was
slightly shallower at large radii, with the mass at the outer boundary of
the plot falling by a factor of $\sim$2. The maximum entropy also falls by
a factor of $\sim$3, agreeing almost exactly with the entropy calculated for
AWM~4. However, neither of these changes significantly alters our
understanding of the behaviour of the system. Gas fraction shows the
largest change, and we show all three gas fraction profiles in
Figure~\ref{fig-Gfrac}. The most obvious change is that when using the
Fukazawa et al. temperature profile for MKW~4, gas fraction is found to
rise outside 0.04$\times$\Rth, producing a maximum value of 1.8\% compared
to $\sim$1\% if our temperature profile is used. This brings it into closer
agreement with AWM~4, but we note that in both versions of the profile,
MKW~4 has a significantly different gas fraction profile from that of AWM~4
between $\sim$0.05 and 0.1$\times$\Rth. We believe this comparison shows
that, with the possible exception of gas fraction, the profiles shown in
Figure~\ref{fig-AWMvsMKW} can be directly compared with confidence in their
accuracy. We therefore proceed with this comparison, but exercise caution
regarding the results for gas fraction.

Several important differences in the structure of the two clusters can be
seen in the profiles. MKW~4 has a steeper gas density profile than AWM~4
(at least over the range covered by these models), with a larger mass of
gas within $\sim$0.02$\times$\Rth. It also has a higher total mass in the
core, though the total mass profiles at larger radii are quite similar. The
gas in the inner region has a significantly shorter cooling time and lower
entropy than is seen in AWM~4. The gas fraction profile suggests that while the two clusters have
comparable gas fractions within $\sim$0.03$\times$\Rth, in AWM~4 the rise
in gas fraction outside this radius is steeper than in MKW~4, and that
AWM~4 may have a considerably higher gas fraction than MKW~4 at large
radii. Gas fraction almost certainly rises to match that of AWM~4 outside
the radius of our analysis; if the Fukazawa et al. temperature profile is used,
the two clusters come into agreement at $\sim$0.2$\times$\Rth; if our
temperature model is used, the turn over in temperature produces a lower
gas fraction, but when the profile levels off at larger radii, as the
\rosat\ results show it does, the fraction will likely rise again.

The similarity in mass outside the core suggests that these two systems are
fundamentally quite similar. The difference in mass in the core is most
likely caused by the difference in mass of the two dominant
ellipticals. Assuming a stellar mass-to-light ratio of 5
$L_\odot/M_\odot$ NGC~4073, the dominant galaxy of MKW~4, has a mass of
5.12$\times$10$^{11}$ \Msol, compared to 2.88$\times$10$^{11}$ \Msol for
NGC~6051. Our X-ray based mass profiles do not exactly match these values
estimated from the optical data, but the difference between the two is
clear.

The other major differences in the profiles can largely be explained as a
consequence of the different temperature profiles of the two clusters. Gas
in the core of MKW~4 has been able to cool, leading to the observed low
entropy and short cooling time. In comparison, gas in the core of AWM~4 has
been heated, lowering the density, flattening the inner entropy profile and
``puffing up'' the halo. The gas fraction profile suggests that the heating
may have effectively moved gas outward, building up a low density core and
a steeper density gradient at larger radii. We cannot determine whether
AWM~4 may have an inherently higher gas fraction than MKW~4 as we do not
have sufficiently deep observations to model the two galaxies out to their
virial radii, but it seems possible that this may be the case to some
degree. Alternatively the two could have similar overall gas fractions at
large radii, but with MKW~4 having a larger core region in which the gas
fraction is low. If the steep rise in the gas fraction profile of AWM~4 is
caused by AGN heating and movement of gas outward, this could indicate
that the process lasted longer and was more effective in MKW~4. However, as
we cannot discriminate between these possibilities with the data available,
we  restrict ourselves to the conclusion that the combination of cooling in
MKW~4, heating in AWM~4 and the difference in central galaxy mass produces
the differences in core gas mass and density.

MKW~4 is probably still in the process of developing a cooling flow, as it
does not yet show any sign of gas cooler than 0.5~keV. If the gas was hotter in
the past, the most likely source of heating is an AGN in NGC~4073, which
has since become quiescent. If this is the case, then the differences we
see between the two clusters are largely the product of the AGN activity in
their dominant galaxies; it happens that we are observing them at different
phases of their duty cycle. In NGC~6051 the AGN has been active for some
time and has heating the surrounding ICM, in the process probably removing
its own fuel source. In NGC~4073, we see the opposite, with the AGN
quiescent for a long period and perhaps soon to reactivate as cooling
provides it with material to accrete. This is a fine example of the
influence of the ICM on the cluster dominant galaxy and vice versa.

We can estimate the energy output required to produce the isothermal
profile observed in AWM~4, assuming that it once had a temperature profile
such as that seen in MKW~4. The total energy required is
$\sim$9$\times$10$^{58}$ erg. If the AGN has an active period of 100 Myr,
this means that a power output of $\sim$3$\times$10$^{43}$ \ergps\ is
required. \citet{Bohringeretal02} estimate the kinetic energy output of the
AGN in M87, NGC~1275 and Hydra~A to be in the range
1.2$\times$10$^{44}$-2$\times$10$^{45}$ \ergps\ based on the bubbles
visible in the ICM of these clusters \citep[see
also][]{McNamara00,Nulsenetal02}. We have estimated the current power
output of the AGN in NGC~6051 from the flux detected from the radio
emission in and around NGC~6051. Radio fluxes have been measured between
26.3 MHz and 10.55 GHz
\citep{Neumannetal94,WrightOtrupcek90,VinerErickson75}, and from these we
calculate a spectral index of -0.9$\pm$0.05, fairly typical for an FR-I
radio galaxy. We estimate the power in this frequency range to be
$\sim$9.4$\times$10$^{40}$ \ergps. However, as \citet{Fabianetal02a}
demonstrate for 3C 84 in the Perseus cluster, this can only be considered a
lower limit on the actual jet power. Based on the cavities observed in the
Perseus cluster core, it is estimated that the power of the jets is
10$^{44-45}$ \ergps, compared to a total emission in the radio band of
10$^{40-41}$ \ergps. Our system is less powerful and we do not see cavities
in the X-ray emission, but is seems reasonable to expect the jet power to
be at least$\sim$10$^{43}$ \ergps.

There are numerous other suggested sources of heating in galaxy clusters,
including star formation and galaxy winds, conduction of heat inward from
the outer halo, shock heating during the merger of a sub-cluster, and AGN
activity. In the case of AWM~4, the first three of these can be dismissed
relatively quickly. Conduction can certainly be ruled out, as the
deprojected temperature profile shows an increasing temperature toward the
core. A recent merger is equally unlikely, as the cluster is relaxed with
no strong substructure. Similarly, there is little evidence of
recent star formation in the central galaxy, and to provide the energy
necessary to heat the halo, star formation would have had to have been
extensive and powerful. Given the strong extended radio source
hosted within NGC~6051, it seems clear that AGN heating is the most likely
cause of the roughly isothermal temperature profile we observe.

\section{Summary and Conclusions}
\label{sec-conc}
AWM~4 is at first glance a good example of a relaxed cluster, with no
significant substructure visible in X-ray images of its halo. Spectral
fitting shows that the cluster core is reasonably well described by a single
temperature plasma model with kT=2.5-2.6 keV and abundance
$\sim$0.5\Zsol. Radial spectral profiles show that the cluster is close to
isothermal out to a radius of $\sim$160 kpc. Abundance rises from
$\sim$0.3\Zsol\ at 160 kpc to $\sim$0.7\Zsol\ in the cluster core. The
highest abundances coincide with the position of NGC~6051 and are
presumably the product of enrichment by supernovae within this galaxy.

Our \xmm\ analysis has revealed some new aspects to the cluster, many of
which appear to be related to the powerful AGN hosted by NGC~6051. The
abundances we observe for the cluster are rather low compared to other
systems which have been observed by \xmm. Possible explanations of this
include mixing of the cluster gas, driven by the activity of the AGN.
Spectral mapping of the cluster reveals a number of features whose
abundances and temperatures diverge from the mean values for the cluster.
The eastern radio jet of 4C+24.36 coincides with a region of low
temperatures which may indicate a cavity in the cluster X-ray halo. The
western jet coincides with the southern edge of a region of high abundance
and high temperature. The high temperatures may be caused by shock heating
of the gas caused as the radio jet compresses the X-ray plasma. However,
the high abundances are more difficult to explain, as is a region of
enhanced temperature and abundance to the south of the galaxy core. These
may indicate enrichment of the ICM through stripping of galaxies in the
cluster, possibly including NGC~6051. The velocity dispersion of the
galaxies in the cluster does not indicate any substructure or outlying
members, but the cluster is elongated along a north-south axis and the
radio source, which appears to have a wide angle tail structure, suggests
motion. Unfortunately we cannot determine motions in the plane of the sky.

Using the best fitting surface brightness profile of the cluster and
assuming it to be approximately isothermal, we have modeled the three
dimensional properties of the cluster X-ray halo and potential well. The
gas mass and total gravitating mass we estimate agree with previous
estimates, both from older X-ray observations and based on the galaxy
population. We find that although the cluster entropy is in the expected
range at 0.1$\times$\Rth, the entropy in the cluster core is somewhat
higher than in comparable systems, and the cooling time is long. The gas
fraction profile also shows a strong increase at 30-70 kpc. We interpret
these factors, in combination with the lack of a cooling core in this
apparently relaxed system, as indications that AGN activity has reheated
the gas in the cluster core. We estimate the power available from the AGN
to be sufficient for this task, and suggest that the current temperature
profile and large-scale radio jets indicate that the AGN has been active
for some time, and may perhaps be nearing the end of its active period. The
lack of ongoing cooling implies that the fuel available for the AGN must
soon run out, leading to dormancy and the reestablishment of cooling in the
inner cluster halo.

\vspace{1cm}
\noindent{\textbf{Acknowledgments}\\
  We are very grateful to S. Helsdon for the use of his 3-d gas properties
  software, and to the referee, Alexis Finoguenov, for a thorough reading
  of the paper and numerous useful suggestions. We would also like to thank
  A. Read and B. Maughan for their advice on XMM analysis, and the use of
  their software.  The authors made use of the NASA and Lyon Extragalactic
  Databases, and were supported in part by NASA grants NAG5-10071 and
  GO2-3186X.

\bibliographystyle{mn2e}
\bibliography{../paper}

\label{lastpage}

\end{document}